\documentclass[11pt]{article}
\usepackage{geometry}                
\usepackage{times,psfrag,graphicx,theorem,epsfig,lscape,amssymb,amsmath,dcolumn,
multirow,epic,float,enumerate,latexsym,ifthen,pifont,comment,bm,subfigure}
\usepackage[round]{natbib}

\newcommand{\blinding}[2]{#1}   

\theoremstyle{plain} 
\theoremstyle{plain} 

\theoremstyle{plain} \newtheorem{ass}{Assumption}
\theoremstyle{plain} 
\theoremstyle{definition} \newtheorem{example}{Example}

\newcommand{\bX}{\mathbf{X}}

\newcommand{\bR}{\mathbf{R}}
\newcommand{\bO}{\mathbf{O}}
\newcommand{\bY}{\mathbf{Y}}

\newcommand{\bZ}{\mathbf{Z}}
\newcommand{\bW}{\mathbf{W}}

\newcommand{\bU}{\mathbf{U}}

\DeclareMathOperator\bE{\mathbb E} 

\def\mis{\textup{mis}}
\def\obs{\textup{obs}}
\def\d{\textup{d}}

\def\sumN{\sum_{i=1}^N}

\def\co{\emph{c}}
\def\nt{\emph{n}}
\def\at{\emph{a}}
\def\df{\emph{d}}

\begin{document}

\begin{center}

{\Large Causal Inference: A Missing Data Perspective}

\blinding{
\medskip
Peng Ding \quad Fan Li
\footnote{Peng Ding is Assistant Professor in Department of Statistics, University of California, Berkeley, CA (email: pengdingpku@berkeley.edu). Fan Li is Associate Professor in Department of Statistical Science, Duke University, Durham, NC (email: fli@stat.duke.edu).
The authors are grateful to Jerry Reiter, the Associate Editor, two reviewers, Avi Feller, Shu Yang, Kari Lock Morgan, Li Ma, and Zhichao Jiang for helpful comments. Peng Ding is partially supported by the IES grant R305D150040, and Fan Li is partially supported by the NSF grant SES-1424688.}

}{}
\end{center}

{\centerline{ABSTRACT}

\noindent Inferring causal effects of treatments is a central goal in many disciplines. The potential outcomes framework is a main statistical approach to causal inference, in which a causal effect is defined as a comparison of the potential outcomes of the same units under different treatment conditions. Because for each unit at most one of the potential outcomes is observed and the rest are missing, causal inference is inherently a missing data problem. Indeed, there is a close analogy in the terminology and the inferential framework between causal inference and missing data. Despite the intrinsic connection between the two subjects, statistical analyses of causal inference and missing data also have marked differences in aims, settings and methods. This article provides a systematic review of causal inference from the missing data perspective. Focusing on ignorable treatment assignment mechanisms, we discuss a wide range of causal inference methods that have analogues in missing data analysis, such as imputation, inverse probability weighting and doubly-robust methods. Under each of the three modes of inference---Frequentist, Bayesian, and Fisherian randomization---we present the general structure of inference for both finite-sample and super-population estimands, and illustrate via specific examples. We identify open questions to motivate more research to bridge the two fields.

\vspace*{0.5cm}
\noindent {\sc Key words}: assignment mechanism, ignorability, imputation, missing data mechanism, observational studies, potential outcome, propensity score, randomization, weighting.
}

\clearpage


\section{Introduction}
Causal inference concerns designs and analyses for evaluating the effects of a treatment. A mainstream statistical framework for causal inference is the potential outcomes framework, under which each unit has a set of potential outcomes corresponding to all possible treatment levels \citep{Neyman23, neyman1935statistical, Rubin74, Rubin77,Rubin78}. Following the \emph{dictum} ``no causation without manipulation'' \citep[][p.235]{Rubin75}, a ``cause'' under the potential outcomes framework strictly refers to a \emph{treatment} or \emph{manipulation}, and a causal effect is defined as a comparison of the potential outcomes under different treatment conditions for the same set of units.

The intrinsic connection to missing data stems from the \emph{fundamental problem of causal inference} \citep{Holland86}, that is, for each unit at most one of the potential outcomes---the one corresponding to the treatment to which the unit is exposed---is observed, and the other potential outcomes are missing. Because the potential outcomes of the same unit are never simultaneously observed, the potential outcomes are often referred to as ``counterfactuals" in the literature. Therefore, causal inference is inherently a missing data problem and estimating causal effects requires properly handling the missing potential outcomes. Despite the intrinsic connection between the two subjects,  statistical analyses of causal inference and missing data---each with a large and sometimes separate literature---also have marked differences in aims, settings and methods. In this paper, we will provide a systematic review of causal inference from a missing data perspective. Below we first go through some primitives.

\subsection{Causal inference and the treatment assignment mechanism}
Consider a simple random sample of units drawn from a target population, indexed by $i\in \{1,...,N\}$, which comprises the participants in a study designed to evaluate the effect of a treatment $W$ on some outcome $Y$.
For example, in comparative effectiveness research, $W$ can be the exposure to a new treatment and $Y$ a health outcome; in economics, $W$ can be the enrollment to a job training program and $Y$ the employment status.
Without loss of generality, we consider binary treatments; extension to general treatment regimes has been discussed elsewhere \citep[e.g.][]{imbens2000role,hirano2004propensity,ImaiDVD04}. Each unit can potentially be assigned to a treatment $w$, with $w=1$ for an active treatment and $w=0$ for control. Let $W_i$ be the binary variable indicating whether unit $i$  is assigned to the treatment ($W_i=1$) or to the control ($W_i=0$). The number of treated and control units are $N_1$ and $N_0$, respectively. At baseline, a vector of $p$ pre-treatment covariates $X_i$ are observed for unit $i$.  From now on, we use bold font to denote matrices or vectors consisting of the corresponding variables for the $N$ units; for example, let $\bW=(W_1,...,W_N)'$ be the $N$-vector of treatment indicators, and $\bX = (X_1',\ldots, X_N')$ be the $N\times p$ covariate matrix. Each unit has a potential outcome under each assignment vector, $Y_i(\bW)$. Assuming the standard stable unit treatment value assumption (SUTVA) \citep{Rubin80}, that is, no interference between units and no different versions of a treatment, each unit has two potential outcomes $Y_i(1)$ and $Y_i(0)$.

The most common causal estimand is the average treatment effect (ATE) -- the difference between the average potential outcomes had all units in a target population were taking the treatment versus not. The ATE has both super-population (PATE) and finite-sample (SATE) versions:
\begin{equation}
\tau^{P}  \equiv \bE\{ Y_i(1)-Y_i(0)\}, \quad \tau^{S} \equiv\frac{1}{N}\sumN\{ Y_i(1)-Y_i(0) \} . \label{eq:ATE}
\end{equation}
In the PATE estimand $\tau^{P}$, all potential outcomes are viewed as random variables drawn from a super-population. In the SATE estimand $\tau^{S}$, all potential outcomes are viewed as fixed values, or, equivalently, all inferences are conditional on the vectors of the potential outcomes $\bY(1)$ and $\bY(0)$. SATE has been mostly discussed in the context of randomized experiments, whereas PATE is usually the target estimand in observational studies. The subtle distinction in their definitions leads to important differences in inferential and computational strategies, as discussed later.  Other estimands of common interest include the average treatment effects for the treated (ATT) and conditional or individual ATE \citep{AtheyImbens15, athey2017estimating}.

Four quantities are associated with each unit $i$, $\{Y_i(0), Y_i(1), W_i, X_i\} $. Only the potential outcome corresponding to the assigned treatment, $Y_i^\obs=Y_i(W_i)$, is observed, and the other potential outcome, $Y_i^\mis=Y_i(1-W_i)$, is missing. Given the observed assignment indicator $W_i$, there is a one-to-one map between $(Y_i^\obs, Y_i^\mis)$ and $\{ Y_i(0),Y_i(1)\} $ with the relationships $Y_i^\obs=Y_i(1)W_i+Y_i(0)(1-W_i)$ and $Y_i^\mis = Y_i(1)(1-W_i) + Y_i(0)W_i$. In general, causal effects are not identifiable without further assumptions. The central identifying assumption concerns the assignment mechanism, that is, the probabilistic process that determines which units receive which treatment condition, and hence which potential outcomes are observed and which are missing. Causal studies can be broadly classified by assignment mechanisms \citep{ImbensRubin15}. The vast majority of causal studies assume an \emph{ignorable assignment mechanism}, also known as \emph{unconfounded} assignment mechanism.

\begin{ass}\label{ass:unconfounded}
\textbf{(Ignorable assignment mechanism)}. An assignment mechanism is ignorable (or unconfounded) conditional on $\bX$ if it does not depend on the potential outcomes:
\begin{eqnarray}
\Pr\left(   \bW \mid  \bY(0), \bY(1), \bX \right)  = \Pr\left(  \bW \mid  \bX \right). \label{am:unconfounded}
\end{eqnarray}
\end{ass}
Assumption \ref{ass:unconfounded} implies that the treatment is randomized among the units with the same value of the observed covariates; in other words, there is no unmeasured confounding. This holds by design in classical randomized experiments where the assignment mechanism is known and depends only on the covariates \citep{Rubin78, Rosenbaum83ps}. However, in observational studies, Assumption \ref{ass:unconfounded} cannot be ensured by design or directly validated by the data; it is an untestable sufficient condition allowing for causal inference.  In this paper, we require ignorable assignment for most of the discussion and will comment on its violation in Section \ref{sec:discussion}.

\subsection{Missing data and the missing data mechanism}\label{sec::missing-mechanism}

The assignment mechanism is a special case of a missing data mechanism. Specifically, in the setting of missing or incomplete data, for each unit $i$, let $O_i$ be the \emph{full data} -- the vector containing all the relevant data, observed or unobserved, and $R_i$ be the missing data indicator with the same dimension as $O_i$. If an entry of $O_i$ is observed, then the corresponding entry of $R_i$ equals $1$; if an entry of $O_i$ is missing, then the corresponding entry of $R_i$ equals $0.$ Let $\bO = (O_1',\ldots, O_N')'$ and $\bR = (R_1', \ldots, R_N')'$ be the collection of the individual data and missing data indicators, respectively. Let $\theta$ and $\psi$ be the set of parameters associated with $\bO$ and $\bR$, respectively; for simplicity,  below we assume that the parameter spaces of $\theta$ and $\psi$ are distinct. Given the missing data indicators, we can partition the relevant data into the observed data $\bO^\obs$ and the missing data $\bO^\mis$. The missing data mechanism---the conditional distribution of $\bR$ given $\bO$ and a parameter $\psi$---is broadly classified into three categories: (i) missing completely at random (MCAR): $\Pr(\bR \mid \bO, \theta, \psi)=\Pr(\bR\mid \psi)$; (ii) missing at random (MAR): $\Pr(\bR \mid \bO, \theta, \psi)=\Pr(\bR \mid \bO^{\obs},\psi)$; (iii) missing not at random (MNAR) $\Pr(\bR \mid \bO, \theta, \psi)=\Pr(\bR\mid \bO^{\obs}, \bO^{\mis}, \psi)$. MCAR is a special case of MAR, and both are ignorable missing data mechanisms \citep{Rubin76, seaman2013meant, mealli2015clarifying}.

If we consider the full data matrix in a causal setting as $\bO = (\bY(1), \bY(0), \bW, \bX)$, then the missing data indicator is $\bR = (\bW, \bm{1}_N -\bW, \bm{1}_{N}, \bm{1}_{N\times p})$; consequently, the observed data contain $\bO^\obs = (\bY^\obs, \bW, \bX)$ and the missing data contain $\bO^\mis = \bY^\mis$. The fundamental problem of causal inference is reflected by the fact that $\bm{1}_N - \bW$, the missing data indicators for $\bY(1)$, and $\bW$, the missing data indicators for $\bY(0)$, sum to $\bm{1}_N$; that is, the two potential outcomes cannot be simultaneously observed and thus missing data are unavoidable. \citet[][p.198]{gelman2014bayesian} call these missing potential outcomes \emph{intentional} missing data, in contrast to the \emph{unintentional} missing data due to unfortunate circumstances such as survey nonresponse, loss to follow-up, censored measurements or clerical errors. Unintentional missing data are extensively discussed in other articles in this issue; in this article we focus on intentional missing data in causal inference, assuming $\bW$ and $\bX$ are fully observed.

\subsection{Connection and distinction} \label{sec::connection}

Based on the representation in Section \ref{sec::missing-mechanism}, the assignment mechanism creates a missing data mechanism of the potential outcomes. There is a broad parallel between the classification of assignment mechanisms in causal inference and the classification of missing data mechanisms. Specifically, the assignment mechanism of completely randomized experiments, namely, $\Pr(  \bW \mid \bY(0),\bY(1), \bX, \theta, \psi)  =\Pr(\bW \mid \psi)$, is in parallel to MCAR. The assignment mechanism of observational studies with unmeasured confounding, namely, $\Pr(  \bW\mid \bY(0), \bY(1), \bX, \theta, \psi ) =\Pr(\bW\mid \bY^{\obs}, \bY^{\mis}, \bX, \psi)$, is in parallel to MNAR. MAR generally corresponds to ignorable assignment mechanisms. However, because the observed covariates, treatment and outcomes play very different roles in causal inference, the original form of ignorable assignment mechanism in Equation \eqref{am:unconfounded} can be generalized to a class of assignment mechanisms by changing the conditioning on the covariates to conditioning on all or a subset of the observed data in the right hand side of Equation \eqref{am:unconfounded}:
\begin{eqnarray}
\Pr\left(  \bW \mid  \bY(0), \bY(1), \bX \right) = \Pr(  \bW \mid  \bY^{\obs},
\bX). \label{am:unconfounded_gen}
\end{eqnarray}
For example, this class includes stratified randomized experiments, the assignment mechanism of which only depends on the covariates, namely, $\Pr( \bW \mid \bY(0),\bY(1), \bX, \theta, \psi)=\Pr(\bW \mid \bX, \psi)$. Another example is the randomized experiments where the units $i=1,\ldots, N$ are assigned sequentially,  with the assignment of unit $i$ depending only on the covariates of units $1,\ldots,i$ and the observed outcomes of units $1,\ldots,i-1$, that is, $\Pr( W_i \mid W_1,\ldots, W_{i-1}, \bY(0), \bY(1), \bX, \theta, \psi )=\Pr(W_i\mid Y^{\obs}_1,\dots, Y^{\obs}_{i-1}, \bX, \psi)$ \citep[][Chapter 3]{ImbensRubin15}. A third example is the sequentially ignoble assignment mechanism in studies with time-varying treatments, where the treatment assignment is ignorable at each time point $t$ conditional on the observed history until time $t-1$, including baseline covariates, observed treatments, and immediate outcomes (i.e. time-varying covariates) \citep{Robins:1986}. Evident from the case of MAR, there is a richer class of assignment mechanisms than missing data mechanisms. In fact, there are a variety of complex assignment mechanisms with no analogous missing data mechanisms, such as latent ignorable assignment with intermediate variables \citep{Frangakis99} and locally ignorable assignment in regression discontinuity designs \citep{li2015evaluating}.

There are several important distinctions between causal inference and the standard missing data analysis. First, the central goal of causal inference is (unconfounded) comparison --- comparing the potential outcomes of the same units under two or more treatment conditions, whereas the common goal in standard missing data analysis is to infer a population parameter---not necessarily comparison---defined on the full data, $\theta=f(\bO)$, from the observed data $\bO^{\obs}$. This has at least two consequences: (a) The crucial concept of covariate balance between two groups in causal inference is usually not considered in missing data analysis; see \cite{zubizarreta2015stable} for an exception; (b) The widely used Fisher randomization test for causal comparisons has no analog in missing data analysis; see Section \ref{sec:bounds} for details. Second, the missingness of potential outcomes is highly structured: for each unit, the missing data indictors of all potential outcomes must sum to one, and thus at least half of the potential outcomes are missing by design. Critically, there is no information to infer the association between the two potential outcomes. In contrast, in standard incomplete data setting, missing data can occur in any part of the full data matrix without structural constraints and usually there is information to infer the association between any two variables. Third, causal inference differentiates between pre-treatment and post-treatment variables, whereas the standard missing data methods do not differentiate between them. For example, it is usually scientifically meaningless to postulate a model of pre-treatment variable conditional on post-treatment variables, e.g. $\Pr(\bX \mid \bY(1), \bY(0))$, but such ordering restriction between variables is rarely imposed in missing data analysis. For example, ordering information between longitudinal variables is usually ignored in the popular multiple imputation by chained equation (MICE) algorithm \citep{van2012flexible}.

Despite these distinctions, the inferential framework of causal inference and missing data analysis are closely related, with common roots in survey sampling. In particular, two overarching methods underpin both causal inference and missing data analysis: weighting and imputation. Although mathematically imputation can be represented as a special form of weighting, these two methods are often derived from different perspectives and implemented differently. Weighting methods weight---usually based on the probability of being missing or assigned to one group---the observed data to represent the full data, and imputation methods physically impute the missing values---often based on a stochastic model---using the observed data. Moreover, sensitivity analysis on the missing data mechanism and the treatment assignment mechanism is routinely conducted in both domains.

Focusing on ignorable assignment mechanisms, in this article we review a wide range of causal inference methods that have analogues in missing data analysis, such as imputation, inverse probability weighting and doubly-robust methods. We also provide examples where ideas originated from causal inference lend to missing data analysis, such as truncation by death and covariate balance. We organize the review by the mode of inference with Frequentist in Section \ref{sec:frequentists}, Bayesian in Section \ref{sec:Bayesian} and Fisherian in Section \ref{sec:Fisherian}. Within each mode of inference, we first present the general structure of causal inference and then illustrate via specific examples; also, depending on the study setting, we switch between two sampling models of the units: we usually, with a few exceptions, adopt the finite-sample model for randomized experiments and the super-population model for observational studies. Section \ref{sec:discussion} concludes with a discussion on open questions and future research directions.


\section{The Frequentist Perspective}
\label{sec:frequentists}

In a broad sense,  the Frequentist or classical inference focuses on repeated sampling evaluation of statistical procedures, such as unbiasedness or consistency, variance and efficiency, and mean squared error of a point estimator, coverage rate of an interval estimator, optimality and minimaxity. Here we discuss the Frequentist perspective in a narrow sense, reviewing the procedures motivated by frequency properties. In particular, we first connect the classical randomization-based inference and model-based imputation in completely randomized experiments, we then review methods using weighting or generally unbiased estimating equations in unconfounded observational studies, and we finally discuss principal stratification in the case of post-treatment variables.

\subsection{Model-based imputation in randomized experiments}\label{sec::freq-FP}
For illustration purpose, we consider a classical completely randomized experiment with covariates, where the inferential goal is to find point and interval estimators for the SATE, $\tau^S$. Recall that in finite-sample inference all potential outcomes $\bY(1)$ and $\bY(0)$ are viewed as fixed and the randomness comes solely from the treatment indicators $\bW$. It is intuitive to impute the missing values---$Y_i(0)$ for the treated units and $Y_i(1)$ for the control units---based on some models of the potential outcomes, and use the imputed potential outcomes, denoted by $\hat{Y}_i(w)$, to estimate the SATE, giving the following \emph{predictive} estimator:
\begin{eqnarray}
\widehat{\tau}^{\text{pre}} = \frac{1}{N} \left\{    \sum_{i=1}^NW_i Y_i(1) + \sum_{i=1}^N (1-W_i) \widehat{Y}_i(1)
-  \sum_{i=1}^NW_i \widehat{Y}_i(0)    - \sum_{i=1}^N (1-W_i) Y_i(0) \right\}.
\label{eq::imputation-formula}
\end{eqnarray}

As a starting point, we consider fitting linear models based on least squares: in the treatment group, $\widehat{Y}_i(1) = X_i' \widehat{\beta}_1 + \widehat{\gamma}_1$, and in the control group, $\widehat{Y}_i(0) = X_i ' \widehat{\beta}_0 + \widehat{\gamma}_0$. Here we highlight that the regression lines include intercepts, and the covariates are centered at zero (i.e. $\bar{X} = \sum_{i=1}^N X_i = 0$).
Define $(\bar{Y}_1^\obs,  \bar{Y}_0^\obs )$ and  $(\bar{X}_1^\obs,  \bar{X}_0^\obs )$ as the means of the outcomes and covariates under treatment and control groups, respectively. Because the least squares coefficients satisfy $  \widehat{\gamma}_w = \bar{Y}_w^\obs -   \widehat{\beta}_w ' \bar{X}_w^\obs \ (w=1,0)$, using some algebra we can simplify the estimator in \eqref{eq::imputation-formula} to
\begin{eqnarray}
\widehat{\tau}^{\text{pre}} =  \widehat{\gamma}_1 -   \widehat{\gamma}_0 =
\left(  \bar{Y}_1^\obs   -    \widehat{\beta}_1' \bar{X}_1^\obs  \right)   - \left(   \bar{Y}_0^\obs   -  \widehat{\beta}_0'  \bar{X}_0^\obs \right)  .
\label{eq::linear-adj}
\end{eqnarray}
Therefore, $\widehat{\tau}^{\text{pre}}$ is identical to the regression adjusted estimator in \citet{lin2013agnostic} motivated by an estimation strategy in survey sampling \citep{cochran2007sampling}. Without using any covariates or simply setting $\widehat{\beta}_1 = \widehat{\beta}_0 = 0$, the estimator $\widehat{\tau}^{\text{pre}}$ reduces to the classical Neymanian unbiased difference-in-means estimator, $\widehat{\tau}^{\text{Neyman}} =   \bar{Y}_1^\obs - \bar{Y}_0^\obs $, which essentially imputes the missing potential outcomes using the observed sample means $\bar{Y}_1^\obs$ and $ \bar{Y}_0^\obs$. \citet{lin2013agnostic} studied the repeated sampling properties of $\widehat{\tau}^{\text{pre}}$ over the distribution of $\bW$, showing that (a) it is consistent and at least as efficient as $\widehat{\tau}^{\text{Neyman}}$ even if the linear potential outcome models are \emph{misspecified}, (b) it is identical to the regression coefficient of $\bW$ in the linear regression $\bY^\obs \sim \bW + \bX + \bW\times \bX$ with full interactions between treatment and covariates, and (c) asymptotically the Huber--White variance estimator \citep{huber1967behavior, white1980heteroskedasticity} is conservative for the randomization-based sampling variance of $\widehat{\tau}^{\text{pre}}$. \citet[][Example 9]{li2017general} supplemented these results with the asymptotic normality and the optimality of $\widehat{\tau}^{\text{pre}}$.

To emphasize, the estimator $\widehat{\tau}^{\text{pre}}$ is motivated by imputing missing potential outcomes based on linear models, but its frequency properties over complete randomization do not require any modeling assumptions. Recently, \citet{bloniarz2016lasso} extended the above approach to deal with high dimensional covariates, replacing the least squares coefficients by the LASSO coefficients \citep{tibshirani1996regression}. Invoking super-population assumptions, \citet{wager2016high} considered other machine learning methods as generalization of the least squares method. So far all these discussions are in the context of completely randomized experiments; rigorous investigations of other types of experiments are desirable.

\subsection{Imputation and weighting in observational studies}
\label{sec::DR}

As the PATE, $\tau^P$, is the primary target estimand in most observational studies, we now shift to the super-population view. We assume the units are a simple random sample from the target population, i.e., $\{Y_i(1), Y_i(0), W_i, X_i\}_{i=1}^N$ are independent and identically distributed. Then, the sufficient condition for identifying $\tau^P$ is strong ignorability \citep{Rosenbaum83ps}, consisting of Assumption \ref{ass:unconfounded} and the overlap (also known as positivity) assumption, which requires the \emph{propensity score}, $ e(X_i) \equiv   \Pr(W_i=1 \mid  X_i)$, is strictly between $0$ and $1$ for all values of $X_i$ .

\subsubsection{Imputation methods}

The most popular method for causal inference in observational studies is regression adjustment \citep{Rubin79}, which, similar to the case in randomized experiments, essentially imputes missing potential outcomes from a regression model.
Let $(p_1, p_0)$ be the proportions of treated and control units. Under Assumption \ref{ass:unconfounded}, the conditional mean functions of the potential outcomes satisfies that for any $w,w',x$,
\begin{eqnarray}
\label{eq::conditional-mean}
m_w(x) \equiv  \bE\{ Y_i(w) \mid  X_i=x  \}
= \bE\{ Y_i(w) \mid  W_i = w',  X_i=x  \}
= \bE\{ Y_i^\obs \mid  W_i = w,  X_i=x  \},
\end{eqnarray}
where the first identity is the definition, the second identity is about the counterfactual mean if $w\neq w'$, and the last identify is the conditional mean function of the observed outcomes. Therefore, the PATE can be identified from the observed data:
\begin{eqnarray}
\tau^P &=& \left[  p_1 \bE\{  Y_i(1) \mid  W_i = 1 \} +  p_0 \bE\{  Y_i(1) \mid  W_i = 0 \} \right ]
- \left[  p_1 \bE\{  Y_i(0) \mid  W_i = 1 \} +  p_0 \bE\{  Y_i(0) \mid  W_i = 0 \} \right ]   \nonumber  \\
&=&   \left[  p_1 \bE\{  Y_i^\obs \mid  W_i = 1 \} +  p_0  \int   m_1(X)  F_0(\d x )  \right ]
- \left[  p_1  \int m_0(x)  F_1(\d x  ) +  p_0 \bE\{  Y_i^\obs \mid  W_i = 0 \} \right ] , \nonumber\\ \label{eq::obs-counterf}
\end{eqnarray}
where $\{ F_1(x)$ and $F_0(x) \}$ are the distribution of the covariates in the treatment and control group, respectively. If $\{  \widehat{m}_1(x), \widehat{m}_0(x)  \}$ are the fitted conditional mean functions based on the treated and control outcome data, then $ \{ \widehat{Y}_i(1), \widehat{Y}_i(0)  \} =   \{  \widehat{m}_1(X_i), \widehat{m}_0(X_i)  \}$ are the fitted values of the possibly missing potential outcomes. This strategy gives the predictive estimator \eqref{eq::imputation-formula}, which reduces to the regression-adjusted estimator \eqref{eq::linear-adj} if both conditional mean functions are modeled as linear.

The predictive estimator \eqref{eq::imputation-formula} also reduces to the popular matching-without-replacement estimator if both  $\widehat{m}_1(X_i)$ and $ \widehat{m}_0(X_i)$ are fitted by the nearest-neighbor regression \citep{abadie2006large}. It is worth commenting more on matching, which has a long tradition in statistics and particularly in causal inference. An early example of matching was given in \cite{chapin1947experimental}. Championed by Rubin (a collection of important papers were reprinted in \cite{rubin2006matched}) and later by Rosenbaum \citep[][Part II]{rosenbaum2010design}, matching has been increasingly embraced by both methodologists and practitioners as a model-free alternative to the regression adjustment method. Instead of a regression model, matching methods use the mean outcome of the units with similar covariates in the opposite group (i.e. matches) to impute the missing potential outcome. In standard missing data analysis, matching-based imputation is also commonly used, where the missing values of one unit are imputed by the observed values of the units who are matched on other characteristics. However, in large complex datasets with incomplete data, model-based imputation methods are generally more flexible and versatile than matching-based ones \citep[e.g.][]{van2012flexible,li2014multiple,murray2016multiple}. Indeed, in these cases matching is often combined with regression models, such as the predictive mean matching (PMM) method \citep{rubin1986statistical,little1988missing}. This is analogous to the bias-corrected matching estimator \citep{AbadieImbens11} in causal inference. In this article we view matching as a special case of imputation methods and refer interested readers to elsewhere for a comprehensive review of the vast literature on matching \citep[e.g.][]{Stuart10,rosenbaum2010design}.

Another representation of the PATE is
\begin{eqnarray}\label{eq::identification-tau}
\tau^P  =   \int  \{   m_1(x)  - m_0(x)  \} F(\d x) = \bE\{  m_1(X) - m_0(X)  \},
\end{eqnarray}
where $F(x)$ is the distribution of the covariates in the whole population. This representation motivates the following {\it projective}  estimator
$$
\widehat{\tau}^{\text{pro}} = \frac{1}{N} \sum_{i=1}^N  \{   \widehat{m}_1(X_i) - \widehat{m}_0(X_i)   \} ,
$$
which is in general different from the {\it predictive} estimator $\widehat{\tau}^{\text{pre}}$, as discussed in the survey sampling literature \citep{firth1998robust}. However, the Normal equation of the linear model and the score equation of the logistic model both satisfy $\sum_{W_i=w}  Y_i^\obs = \sum_{W_i=w}  \widehat{m}_w(X_i) $, ensuring that $\widehat{\tau}^{\text{pro}} = \widehat{\tau}^{\text{pre}}$ if we fit separate linear or logistic models with intercepts for the observed outcomes under treatment and control. In contrast to the case of completely randomized experiments as discussed in Section \ref{sec::freq-FP}, consistency of the estimators $\widehat{\tau}^{\text{pro}} $ and $\widehat{\tau}^{\text{pre}}$ in observational studies does rely on correct specification of the conditional mean functions $m_1(x)$ and $m_0(x)$ \citep{tsiatis2008covariate}.

A special case of the regression adjustment/imputation method is to use the estimated propensity scores as a predictor additional to the covariates in the outcome model; such a model has been shown in many empirical studies to outperform the same model without the propensity scores. In particular, \citet{little2004robust} and \citet{zhang2009extensions} advocated to use penalized splines of the estimated propensity scores in the outcome model, in the context of both causal inference and missing data. The common idea originates from survey methodology, where the sampling probabilities (equivalently the survey weights) usually contain information about the units besides the covariates and have been used to augment the inference about population parameters.

\subsubsection{Weighting methods}

A second class of widely-used causal inference methods in observational studies is weighting, particularly propensity score weighting. \citet{Rosenbaum83ps} showed that if the assignment mechanism is unconfounded given the covariates $X_i$ (i.e. Assumption \ref{ass:unconfounded}), then it is also unconfounded given the scalar propensity score $e(X_i)$. The classical Horvitz--Thompson estimator \citep{HT52} in survey sampling motivates the following representation of the PATE:
\begin{eqnarray}
\label{eq::inverse-prob-weight}
\tau^P = \bE  \left\{   \frac{  W_i Y_i^\obs }{  e(X_i) }   \right\}   -  \bE  \left\{   \frac{  (1-W_i)  Y_i^\obs }{  1 - e(X_i) }   \right\}   .
\end{eqnarray}
This suggests that one can define an inverse probability weight (IPW) for each unit in the treatment as $1/e(X_i)$ and in the control group as $1/\{1-e(X_i)\}$, and estimate the PATE by the difference in the weighted average of the outcomes in the two groups. \cite{LiLockZaslavsky16} show that the IPW is a special case of the general class of \emph{balancing weights}, which balance the weighted distributions of the covariates between treatment and control groups in any pre-specified target population.

If we fit a model for $\Pr(W_i=1 \mid  X_i)$ and obtain the estimated propensity score $\widehat{e}(X_i)$, then the sample analogue of \eqref{eq::inverse-prob-weight} gives the IPW estimator of the PATE:
\begin{eqnarray}
\widehat{\tau}^{\text{ipw}} =  \frac{1}{N} \sum_{i=1}^N  \frac{  W_i Y_i^\obs }{  \widehat{e}(X_i) }
- \frac{1}{N} \sum_{i=1}^N   \frac{  (1-W_i)  Y_i^\obs }{  1 - \widehat{e}(X_i) } .
\label{eq::ipw}
\end{eqnarray}
Note that the consistency of $\widehat{\tau}^{\text{ipw}} $ requires a correctly specified propensity score model.
If we replace the $N$'s in the denominators by $\sum_{i=1}^N W_i /   \widehat{e}(X_i)  $ and $\sum_{i=1}^N (1-W_i) /  \{ 1-  \widehat{e}(X_i) \} $, respectively, the resulting estimator is called the Hajek estimator, which often has smaller mean squared errors \citep{hajek1971comment}.
When the fitted values $ \widehat{e}(X_i)$ are close to $0$ or $1$, the IPW estimator is unstable, leading to large variance and bias. Common remedies include trimming units with extreme weights \citep{crump2009dealing}, coarsening propensity scores \citep{zhou2015coarsened}. Another approach is to stratify on $\widehat{e}(X_i)$ \citep{Rosenbaum83ps}, which can be viewed as a special case of $\widehat{\tau}^{\text{ipw}} $ with the estimated propensity scores coarsened before being plugged into \eqref{eq::ipw}. \cite{LiLockZaslavsky16} advocated to change the inverse probability weights in $\widehat{\tau}^{\text{ipw}}$ to the overlap weights (i.e., weights control units by $e(X_i)$ and treated units by $1-e(X_i)$), which shift the focus to the target population with the most overlap and automatically bypass the problem of extreme weights.

The Horvitz--Thompson idea that a unit being weighted inverse-proportionally to its probability of being sampled applies to both causal inference and missing data. Unsurprisingly, the IPW approach is also widely adopted in missing data analysis, particularly for handling incomplete outcome data. For example, in order to estimate a population parameter from a sample with incomplete outcome data, we can first estimate the probability of being observed based on the covariates for each unit, corresponding to the propensity score, and then run a weighted regression model where each observed unit is inversely weighted by its probability of being observed.

\subsubsection{Combining imputation and weighting: Doubly-robust methods}

A third class of popular methods is the doubly-robust estimation, which combines the virtues of regression adjustment and propensity score weighting. Based on \eqref{eq::identification-tau} and \eqref{eq::inverse-prob-weight}, it is straightforward to verify that
\begin{eqnarray*}
\tau^P &=&  \bE\left\{   \frac{ W_i Y_i^\obs   }{  e(X_i) }   - \frac{W_i - e(X_i)}{  e(X_i) } m_1(X_i)      \right\}
-  \bE\left\{     \frac{ (1-W_i) Y_i^\obs  }{  1- e(X_i) }    + \frac{ W_i - e(X_i)   }{ 1- e(X_i)}  m_0(X_i) \right\} \\
&=& \bE\left[    m_1(X_i) + \frac{    W_i\{  Y_i^\obs  - m_1(X_i)  \}    }{  e(X_i) }   \right]
- \bE\left[   m_0(X_i) +    \frac{   (1- W_i)  \{  Y_i^\obs  - m_0(X_i)  \}    }{  1 - e(X_i) }      \right].
\end{eqnarray*}
This motivates the following estimator
{\small
\begin{eqnarray}
\widehat{\tau}^{\text{dr}} &=&  \frac{1}{N} \sum_{i=1}^N  \left\{  \frac{ W_i Y_i^\obs   }{  \widehat{e}(X_i) }   - \frac{W_i -  \widehat{e}(X_i)}{   \widehat{e}(X_i) } \widehat{m}_1(X_i) \right\} -  \frac{1}{N} \sum_{i=1}^N  \left\{     \frac{ (1-W_i) Y_i^\obs  }{  1-  \widehat{e}(X_i) }    + \frac{ W_i -  \widehat{e}(X_i)   }{ 1-  \widehat{e}(X_i)}  \widehat{m}_0(X_i) \right\} \label{eq::ipw-reg} \\
&=&  \frac{1}{N} \sum_{i=1}^N  \left[    \widehat{m}_1(X_i) + \frac{    W_i\{  Y_i^\obs  - \widehat{m}_1(X_i)  \}    }{   \widehat{e}(X_i) }   \right]
-  \frac{1}{N} \sum_{i=1}^N  \left[    \widehat{m}_0(X_i) +    \frac{   (1- W_i)  \{  Y_i^\obs  - \widehat{m}_0(X_i)  \}    }{  1 -  \widehat{e}(X_i) }      \right] .  \label{eq::reg-ipw}
\end{eqnarray}
}
The equivalent forms \eqref{eq::ipw-reg} and \eqref{eq::reg-ipw} follow from trivial mathematical manipulations but have profound statistical implications. We can view \eqref{eq::ipw-reg} as the IPW estimator augmented by an outcome model, and show that if the propensity score model is correctly specified then $\widehat{\tau}^{\text{dr}} $ is consistent for $\tau$ no matter whether the outcome models are correctly specified. Alternatively, we can view \eqref{eq::reg-ipw} as the projective estimator augmented by inverse probability weighting, and show that if the outcome models are correctly specified then $\widehat{\tau}^{\text{dr}} $ is consistent for $\tau$ no matter whether the propensity score model is correctly specified. This property is called double-robustness (DR) \citep{Scharfstein99, Bang05}, that is, $\widehat{\tau}^{\text{dr}} $ is consistent for $\tau$ if either the propensity score model or the outcome models are correctly specified.
Moreover, the DR estimator achieves the semiparametric efficiency bound if both the propensity score and outcome models are correctly specified \citep{robins1995analysis, robins1997toward}. For more discussion on efficiency, see \citet{Hahn98}, \citet{hirano2003efficient}, \citet{Imbens04} and \citet{qin2017biased}. The DR estimator is closely connected to missing data analysis; in fact, the concept of DR was first proposed in the context of missing data and was later formally extended to causal inference \citep{Lunceford04, Bang05}.

The DR estimator has attracted increasing attention from both theoreticians and practitioners. Below we comment on several issues appearing frequently in the recent literature. First, the propensity score model is often chosen to be logistic in practice, and the outcome models linear. With low dimensional covariates, we can add power series of the covariates to approximate complex propensity score and outcome models \citep{newey1997convergence, hirano2003efficient, mercatanti14debit}. Second, with high dimensional covariates, the DR estimator works well under more general assumptions than many other estimators, and in order to achieve better asymptotic properties, for example, $\sqrt{n}$-consistency, the DR estimator is probably the only way to go \citep{van2011targeted, belloni2014inference, chernozhukov2016double, athey2016approximate,belloni2017program}. More importantly, in high dimensions, it is crucial to regularize the estimators as suggested by, for example, \citet{chernozhukov2016double} and \citet{athey2017estimating}. Third,
the unstable properties of $\widehat{\tau}^{\text{dr}}$ can be more severe than $\widehat{\tau}^{\text{ipw}}$ with extreme estimated propensity scores, especially when both the propensity score and outcomes models are misspecified \citep{kang2007demystifying}. In this case, trimming or truncating the estimated propensity scores seems crucial for reliable analysis. Alternatively, \citet{graham2012inverse}, \citet{hainmueller2012entropy} and \citet{imai2014covariate} proposed to construct weights that balance the covariates directly, which can avoid the problem of extreme estimated propensity scores. Interestingly, the idea of building propensity score models based on covariate balance has long been used in the causal inference, but it is only recently adopted in missing data problems. Specifically, \cite{zubizarreta2015stable} constructed more stable weighting estimators for missing data analysis by directly constructing weights to balance covariates of units with observed and unobserved outcomes.

\subsection{Principal stratification} \label{sec::freq-PS}

A important concept in causal inference is post-treatment variables -- variables that are potentially affected by the treatment and also affect the outcome. \citet{cochran1957analysis} and \cite{Rosenbaum84} shows that adjusting for post-treatment variables in the same fashion as adjusting for pre-treatment covariates would bias the estimation of causal effects in general. Since the landmark papers by \cite{Angrist96} and \cite{FrangakisRubin02}, a large literature on this topic has been developed. The post-treatment variable settings include a wide range of specific examples, such as treatment noncompliance \citep[e.g.][]{ImbensRubin97}, outcomes truncated by ``death'' \citep[e.g.][]{Rubin06, Zhang09}, surrogate endpoints \citep[e.g.][]{Gilbert08, ZiglerBelin12}, mediation analysis \citep[e.g.][]{VanderWeele2008, Gallop09,Elliott10}, and fuzzy regression discontinuity designs \citep[e.g.][]{li2015evaluating,ChibJacobi15}.

\subsubsection{Randomized experiments with noncompliance}
To introduce the basic setup, we start with the simplest setting of a completely randomized experiment with noncompliance.  We need some new notation to accommodate the post-treatment setting. For unit $i\ (i=1,\ldots, N)$ in a study sample, let $Z_i$ be the treatment assigned to ($1$ for treatment and $0$ for control), and $W_i^{\obs}$ be the treatment received ($1$ for treatment and $0$ for control). When $Z_i \neq W_i^{\obs}$, noncompliance occurs. Because $W_i$ is a post-treatment variable, it has two potential outcomes, $W_i(0)$ and $W_i(1)$, with $W_i^{\obs}=W_i(Z_i)$. As before, the outcome $Y_i$ also has two potential outcomes, $Y_i(0)$ and $Y_i(1)$. \citet{Angrist96} classified the units according to the joint potential treatment statuses $U_i = (W_i(1), W_i(0))$, which was later called the principal stratification by \citet{FrangakisRubin02}. The principal stratum $U_i$ takes four values: $(1,0) = \co$ for compliers, $(0,0) = \nt$ for never takers, and $(1,1) =\at$ for always takers, and $(0,1) = \df $ for defiers. Due to the fundamental problem of causal inference, individual principal stratum $U_i$ are not observed. Indeed, the observed cells of $Z$ and $W$ usually consists of units from more than one principal strata.

The key property of principal strata is that they are, by definition,
not affected by the treatment assignment, and thus can be regarded
as a pre-treatment variable. Therefore, comparisons of $Y_i(1)$ and
$Y_i(0)$ within a principal stratum---the principal causal effects (PCEs)---have a causal interpretation:
$$
\tau_u = \bE\{ Y_i(1) - Y_i(0) \mid U_i  =u \} = \bE\{ Y_i(1)   \mid U_i  =u \} - \bE\{  Y_i(0) \mid U_i  =u\},
$$
for $u=n,c,a,d$. Note that PCE also has both finite-sample and super-population versions. The conventional causal estimand in the noncompliance setting is the intention-to-treat (ITT) effect, ignoring noncompliance; ITT can be decomposed into the sum of the four PCEs:
$$
\tau^P = \bE\{ Y_i(1) - Y_i(0) \} = \sum_{u=\co, \nt, \at, \df}   \pi_u \tau_u^P,
$$
where $\pi_u  = \Pr( U_i = u) $ is the proportion of the stratum $u$. Inferences about the PCEs are often scientifically relevant; for example, the PCE for compliers---commonly known as the complier average causal effect (CACE) or the local average treatment effect (LATE) \citep{ImbensAngrist94}---is informative about the efficacy of the treatment received, whereas the ITT is informative about the effectiveness of the treatment assigned.

The main challenge to the inference of principal stratification is the missing latent stratum membership $U_i$. Specifically, the observed cells of $Z$ and $W^{\obs}$ usually consists of a mixture of units from more than one principal strata, and one has to disentangle the causal effects for different principal strata from observed data. Therefore, the general structure of inference with principal stratification is similar to that of a mixture model, and standard methods for mixture models such as the EM algorithm and data augmentation are routinely used in principal stratification, as illustrated below in Section \ref{sec::em}.

\subsubsection{Implicit weighting via the EM algorithm}
\label{sec::em}

Throughout the discussion, we will use $\Pr(\cdot\mid \cdot)$ and $\theta_{\cdot|\cdot}$ to denote generic conditional distributions and the corresponding parameters, respectively.
If we have full parametric models for $\Pr(U_i\mid X_i, \theta_{W|X})$ and $\Pr(  Y_i(z) \mid U_i, X_i, \theta_{Y|U,X}  ) $, then we can use the EM algorithm to obtain the maximum likelihood estimator for $\theta = (\theta_{W|X}, \theta_{Y|U,X})$ treating $\bU$ as the missing data.
Introducing the shorthand notation $\pi_{i,u}=\Pr( U_i=u \mid X_i, \theta_{W|X})$ and $f_{i,uz}=\Pr(Y_i(z) \mid U_i=u, X_i, \theta_{Y|U,X})$, the complete-data log-likelihood, based on $(\bZ,  \bW^\obs, \bU, \bY^\obs, \bX)$, is
\begin{eqnarray*}
l(\theta ,  \bU)
&=&
\sum_{Z_i=1, W_i^\obs = 1}   I(U_i = \co ) \left\{  \log \pi_{i,\co} + \log f_{i,\co 1}  \right\} + \sum_{Z_i=1, W_i^\obs = 1}  I(U_i = \at ) \left\{  \log \pi_{i,\at} + \log f_{i,\at 1}  \right\}  \\
&&+
\sum_{Z_i=1, W_i^\obs = 0}  I(U_i = \nt )\left\{  \log \pi_{i,\nt} + \log  f_{i,\nt 1} \right\} + \sum_{Z_i=1, W_i^\obs = 0}  I(U_i = \df )  \left\{  \log \pi_{i,\df} + \log f_{i, \df 1}  \right\} \\
&&+
\sum_{Z_i=0, W_i^\obs = 1}  I(U_i = \at ) \left\{ \log \pi_{i,\at} + \log   f_{i,\nt 0}  \right\}  + \sum_{Z_i=0, W_i^\obs = 1}  I(U_i = \df ) \left\{  \log \pi_{i,\df} + \log f_{i, \df 0}  \right\}  \\
&&+
\sum_{Z_i=0, W_i^\obs = 0}  I(U_i = \co )\left\{  \log \pi_{i,\co} + \log   f_{i,\co 0}  \right\} + \sum_{Z_i=0, W_i^\obs = 0}  I(U_i = \nt ) \left\{  \log \pi_{i,\nt} + \log f_{i, \nt 0} \right\}.
\end{eqnarray*}
We emphasize that $l(\theta ,  \bU)$ is a function of the parameter $\theta$ and depends on the missing principal strata $\bU.$

In the E-step of the EM algorithm, we need to find $Q(\theta\mid \theta^{[t]}) = \bE\{ l(\theta ,  \bU)  \mid  \bZ,  \bW^\obs,  \bY^\obs, \bX , \theta^{[t]}  \}$, the conditional expectation of the complete-data log-likelihood given the observed data and the value of the parameter at iteration $t$, which reduces to calculating the conditional probabilities of $ \Pr^{[t]}(U_i = u) =  \Pr(U_i = u\mid Z_i, W_i^\obs, Y_i^\obs, X_i, \theta^{[t]} )$. For example, for unit with $Z_i=1$ and $W_i^\obs = 1$, we have
$$
\text{Pr}^{[t]}(U_i = \co)  =  \frac{  \pi_{i,\co}^{[t]}  f_{i,\co 1}^{[t]}     }{     \pi_{i,\co}^{[t]}  f_{i,\co 1}^{[t]} +  \pi_{i,\at}^{[t]}  f_{i,\at 1}^{[t]}   },\quad
\text{Pr}^{[t]}(U_i = \at)  =  1- \text{Pr}^{[t]}(U_i = \co) ,
$$
where $\pi_{i,u}^{[t]} $ and $ f_{i,u 1}^{[t]} $ are evaluated at $\theta^{[t]}$. This effectively creates two weighted observations: one has $(Z_i=1,W_i^\obs=1,U_i=\co,Y_i^\obs,X_i)$ with weight $\Pr^{[t]}(U_i = \co) $, and the other has $(Z_i=1,W_i^\obs=1,U_i=\at,Y_i^\obs,X_i)$ with weight $\Pr^{[t]}(U_i = \at) $.
For units with other combinations of $Z_i$ and $W_i^\obs$, we can similarly obtain $ \Pr^{[t]}(U_i = u)$ and create weighted observations.

In the M-step of the the EM algorithm, we need to maximize $Q(\theta\mid \theta^{[t]})$, or, equivalently, the log likelihood function from the weighted samples obtained from the E-step. To be more specific, \citet{Zhang09} and  \citet{frumento2012evaluating} proposed to use a multinomial logistic model for $\Pr(U_i \mid X_i, \theta_{W|X})$ and normal linear models for $\Pr( Y_i(z) \mid U_i, X_i, \theta_{Y|U,X} ) $. If the parameter spaces of $\theta_{W|X}$ and $\theta_{Y|U,X} $ are distinct, then we can update them in the M-step separately: based on the weighted samples, $\theta_{W|X}^{[t+1]}$ can be obtained by fitting a weighted multinomial logistic regression, and $\theta_{Y|U,X}^{[t+1]}$ can be obtained by fitting weighted least squares with different combinations of $(Z_i=z, U_i=u)$.

In the analysis of randomized experiments with noncompliance, two assumptions are often invoked: monotonicity (i.e. $\Pr(U_i=d)=0$), and exclusion restriction for noncompliers (i.e. $Y_i(1)=Y_i(0)$ for $U_i=a$ and $n.$) These assumptions ensure nonparametric identification of the CACE, and thus the result is not sensitive to parametric assumptions. Without these two assumptions, subtle identification and inferential issues arise.
Although the identifiability of parametric models is guaranteed by the identifiability of mixture models, the likelihood function may display pathological behavior \citep{ding2011identifiability, frumento2016fragility, feller2016principal}, and standard Frequentist inferential tools, for example the bootstrap, do not apply. Without these two assumptions, the PCEs are only \emph{partially} identified without additional parametric assumptions; the identification issue will be further discussed in the Bayesian perspective \ref{sec:Bayes_FP}.

\subsubsection{Truncation by death or missing data?}
Though post-treatment variables do not have direct analogue in missing data analysis, principal stratification has several applications that shed new lights to some challenging missing data problems. One such application is the truncation by ``death'' problem. Consider a medical study that aims to evaluate the effect of a new treatment ($Z_i$) on the quality of life ($Y_i$); here some patients may die before the outcome was measured and the post-treatment variable $W_i$ is the survival status. Another example in labor economics is to evaluate the effects of a job training program ($Z_i$) on hourly wage ($Y_i$); here some subjects may be unemployed and $W_i$ is the employment status. A common feature of these two problems is that the outcome is not well-defined or truncated if $W_i=0$, i.e., unit $i$ is dead or unemployed. The name ``truncation by death'' comes from the original medical application \citep{Rubin06}.

Traditionally, the truncation by ``death'' problem was viewed as a standard missing data problem, with $W_i$ being the missing data indicator. A famous model in this context is the Heckman selection model \citep{heckman1979sample}, which consists of an outcome equation ($\bY\sim \bZ + \bX$) and a selection equation ($\bW \sim \bZ + \bX$). This effectively assumes that the outcome is well-defined for all units, which, however, is not easy to justify scientifically. \citet{Rubin06} proposed to tackle this problem from a principal stratification perspective, namely, estimating the average causal effect among the units who would always survive under both treatment and control---the survivor average causal effect $\tau^{\text{SACE}}$:
$$
\tau^\text{SACE} = \bE\{ Y_i(1) - Y_i(0) \mid W_i(1) = W_i(0) = 1 \}.
$$
Estimating $\tau^{\text{SACE}}$ is again a missing data problem, but now the missingness arises from the latent stratum labels $\bU$ rather than the original outcome $\bY$.  This alternative view is arguably more scientifically relevant than the traditional one. \citet{Zhang09} developed the EM-based estimation strategy in Section \ref{sec::em}; recent extensions can be found in \citet{ding2011identifiability}, \citet{frumento2012evaluating}, \citet{yang2016using} and \citet{ding2016principal}.

\section{The Bayesian Perspective}
\label{sec:Bayesian}

The Bayesian paradigm offers a unified inferential framework for missing data and causal inference. Under the potential outcomes framework, each unit is associated with several quantities including the potential outcomes, some of which are observed and some are missing. Bayesian inference considers the observed values of these quantities to be realizations of random variables and the unobserved values to be unobserved random variables \citep{Rubin78}, which are no different from unknown model parameters. Because inferences for the finite-sample and super-population estimands have subtle but important differences, below we will discuss them separately.  

\subsection{Finite-sample inference} \label{sec:Bayes_FP}
Finite-sample causal estimands, defined as functions of the $N\times 2$ potential outcome matrix $\{  \bY(0), \bY(1) \}$, can also be represented as functions of missing and observed potential outcomes, because $\tau^{S} =\tau(\bY(0), \bY(1)) =\tau(\bY^{\obs}, \bY^{\mis}, \bW)$. Finite-sample Bayesian causal inference centers around building a model to impute the missing potential outcomes given the observed data and then deriving the posterior distributions of the estimands.


Let $\Pr\{  \bY(0), \bY(1), \bW, \bX \mid   \theta \}$  be the joint probability density function of the random variables for all units governed by a model parameter $\theta$. For ease of exposition, we assume these random variables for each unit are i.i.d. conditional on $\theta$ and thus we can factor the complete-data likelihood as
$
 \prod_{i=1}^N  \Pr\{   Y_i(0), Y_i(1), W_i, X_i \mid  \theta   \}.
$
Imposing a prior distribution $p(\theta)$ on the parameter, we obtain the joint posterior distribution of the missing data and the parameter as
\begin{eqnarray}
\Pr(   \bY^\mis, \theta \mid  \bY^\obs, \bW, \bX  )
\propto   p(\theta)  \prod_{i=1}^N  \Pr(   Y_i(0), Y_i(1), W_i, X_i \mid  \theta  )   .
\label{eq::joint-posterior}
\end{eqnarray}
Note that in Equation \eqref{eq::joint-posterior} and the formulas below, we follow the convention of Bayesian statistics and use $\propto$ to denote ``proportional to'' a probability density, dropping the normalizing constant. Formula \eqref{eq::joint-posterior}---followed immediately from the Bayes' Theorem---is fundamental to finite-sample  Bayesian causal inference: One can first obtain the posterior distribution of $\bY^\mis$ from \eqref{eq::joint-posterior}, and then obtain the posterior distribution of $ \tau^{S} =\tau(\bY^{\obs}, \bY^{\mis}, \bW)$ because $\bY^\obs$ and $\bW$ are known.

To make the role of \eqref{eq::joint-posterior} more explicit, we discuss two strategies to simulate the posterior distribution of $\bY^\mis$, motivated by both computational and statistical considerations. The first strategy is based on data augmentation \citep{Tanner87} or more generally Gibbs sampling \citep{Gelfand90}, that is, we iteratively simulate $\bY^\mis$ and $\theta$ given each other and the observed data, based on $\Pr(\bY^{\mis}\mid \bY^{\obs}, \bW, \bX, \theta)$ and $\Pr(\theta\mid\bY^{\mis}, \bY^{\obs}, \bW, \bX)$. Although this is a routine algorithm in Bayesian inference, explicating some steps is helpful for gaining insights. In particular, given the observed data and the parameter $\theta_{Y|X}$, the missing potential outcomes $\bY^\mis$ have a posterior distribution as
{\small
\begin{eqnarray}
 \Pr(\bY^{\mis}\mid \bY^{\obs}, \bW, \bX, \theta)
&\propto &\Pr(\bY(0),\bY(1), \bW, \bX \mid  \theta)  \nonumber \\
&\propto & \prod_{i=1}^N \Pr(W_i \mid Y_i(0),Y_i(1), X_i, \theta) \Pr(Y_i(0),Y_i(1)\mid X_i, \theta)\Pr(\bX_i \mid \theta) .\label{AM}
\end{eqnarray}
}
It is common to parameterize the assignment mechanism, the distribution of potential outcomes, and the distribution of covariates with different and \emph{a priori} independent sets of parameters, say $\theta_{W|X}$, $\theta_{Y|X}$ and $\theta_{X}$, respectively. Then, given unconfoundedness, the assignment mechanism $\Pr(W_i\mid Y_i(0),Y_i(1),\bX_i,\theta)$ and the covariate distribution $\Pr(X_i\mid\theta)$ drop out in \eqref{AM}, which simplifies to:
\begin{eqnarray} \label{AM_reduced}
&&\Pr(\bY^{\mis}\mid\bY^{\obs}, \bW, \bX, \theta) \nonumber\\
&\propto& \prod_{i: W_i=1} \Pr(Y_i(0),Y_i(1)\mid X_i, \theta_{Y|X})    \prod_{i: W_i=0} \Pr(Y_i(0),Y_i(1)\mid X_i, \theta_{Y|X})  \nonumber\\
&\propto &    \prod_{i: W_i=1}   \Pr(   Y_i(0) \mid  Y_i(1),  X_i, \theta_{Y|X}  )
 \prod_{i: W_i=0}      \Pr(  Y_i(1) \mid  Y_i(0),  X_i, \theta_{Y|X} ).
 \label{eq::impute-missing}
\end{eqnarray}
Because the posterior distribution \eqref{eq::impute-missing} factors into $N$ terms, for treated units we impute the missing control potential outcomes from $\Pr(   Y_i(0) \mid  Y_i(1),  X_i, \theta_{Y|X} ) $, and for control units we impute the missing treatment potential outcomes from $ \Pr(   Y_i(1) \mid  Y_i(0),  X_i, \theta_{Y|X} ) .$ Imputing the missing potential outcomes $\bY^\mis$ depends crucially on the joint distribution of $\{ Y_i(1), Y_i(0) \}$ given $X_i$. Given the observed data and the imputed $\bY^\mis$, the posterior distribution of $\theta_{Y|X}$ can be obtained by a complete-data analysis based on $\Pr( \theta_{Y|X} \mid   \bY(1), \bY(0) , \bX ) \propto p(\theta_{Y|X}) \prod_{i=1}^N  \Pr(Y_i(1), Y_i(0) \mid  X_i, \theta_{Y|X} )   $.

The above general framework was first proposed by \citet{Rubin78} and has been widely adopted in the literature; recently \citet{Heckman14} extended it to more specific econometric models. However, this strategy has a limitation of mixing fully identifiable and non (or weakly) identifiable parameters. Due to the simultaneous presentation of several inferential frameworks in this article, we need to clarify the different notions of identifiability in each framework. Under the Frequentist paradigm, a parameter is identifiable if it can be expressed as a function of the distribution of the observed data \citep{bickel2015mathematical}. In other words, a parameter is identifiable in the Frequentist sense if two distinct values of it give two different distributions of the observed data. Under the Bayesian paradigm, there is no consensus. For example, \citet{lindley1972bayesian} argued that in Bayesian analysis, all parameters are identifiable because with proper prior distributions, posterior distributions are always proper. However, \citet{gustafson2015bayesian} argued that a parameter is only \emph{weakly} or \emph{partially} identifiable, if a substantial region of its posterior distribution is flat, or its posterior distribution depends crucially on its prior distribution even with large samples.
Specifically, because $Y_i(1)$ and $Y_i(0)$ are never jointly observed, the data provide little information about the parameter that governs the association between $Y_i(1)$ and $Y_i(0)$. Consequently, the posterior distribution of $\tau^{S}$ will be sensitive to its prior distribution. Therefore, it may be more sensible to isolate the parameter---denoted by $\theta^\text{m}$---that governs the marginal distributions from the parameter---denoted by $\theta^\text{a}$--- that governs the association between $Y_i(1)$ and $Y_i(0)$. This motivates the following strategy to simulate the posterior distribution of $\bY^\mis$.

The second strategy is based on the definition of conditional probability, that is,
$$\Pr(\bY^{\mis}, \theta \mid \bO^{\obs})= \Pr(\theta \mid \bO^{\obs})\Pr(\bY^{\mis}\mid \theta, \bO^{\obs}),$$
where $\bO^\obs = (\bX, \bY^\obs, \bW)$. Here we first simulate $\theta$ given the observed data, $\Pr(\theta \mid \bO^{\obs})$, and then simulate $\bY^\mis$ given $\theta$ and the observed data, $\Pr(\bY^{\mis}\mid \theta, \bO^{\obs})$. Following the above arguments, we partition $\theta_{Y|X}$ into $\theta_{Y|X}^\text{m}$ and $\theta_{Y|X}^\text{a}$, and impose independent priors on them. Therefore, the posterior distribution of $\theta$ becomes
\begin{eqnarray}\label{eq:theta_post}
\Pr(\theta \mid \bY^\obs , \bW, \bX ) \propto   p(  \theta_{Y|X}^\text{a} )    p( \theta_{Y|X}^\text{m} )
\prod_{W_i=1}   \Pr(   Y_i(1) \mid X_i,   \theta_{Y|X}^\text{m}  )  \prod_{W_i=0}   \Pr(  Y_i(0) \mid X_i,   \theta_{Y|X}^\text{m}  ) .
\end{eqnarray}
Not surprisingly, the posterior distribution of $\theta_{Y|X}^\text{m}$ is updated by the likelihood, but the posterior of $\theta_{Y|X}^\text{a}$ remains the same as its prior. This is due to the \emph{a priori} independence between $ \theta_{Y|X}^\text{m}$ and $\theta_{Y|X}^\text{a}$; otherwise the posterior of $\theta_{Y|X}^\text{a}$ will be updated {\it indirectly} via its dependence on $ \theta_{Y|X}^\text{m}$, as pointed out by \citet{Gustafson09}. \citet{richardson2010transparent} suggested transparent parametrization by separating identifiable and non-identifiable parameters in Bayesian causal inference;  \citet{ding2016potential} suggested sensitivity analysis by varying $ \theta_{Y|X}^\text{a}$ in a certain range (more details in Example 2 below). After drawing $\theta_{Y|X}$ from its posterior distribution \eqref{eq:theta_post}, we can impute the missing potential outcomes $\bY^\mis$ using the same formula as \eqref{eq::impute-missing}.

To illustrate the idea, we discuss two examples arising from the Bayesian analysis of completely randomized experiments.

\begin{example}\normalfont
[Two-by-two table]
In a completely randomized experiment with treatment $W$ and binary outcome $Y$, the joint potential outcomes $\{ Y_i(1), Y_i(0) \}$ can take four values with counts $N_{y_1y_0} = \#\{  i: Y_i(1)  = y_1, Y_i(0) = y_0 \}$ for $y_1,y_0=0,1$, and the observed data can be summarized by a two-by-two table with counts $n_{wy} = \#\{ i: W_i = w, Y_i^\obs = y \}$ for $w,y=0,1$.

Following the first strategy, we can impose a multinomial model for $\{ Y_i(1), Y_i(0)\}$ with a Dirichlet prior for the probabilities:
\begin{eqnarray*}
\{ Y_i(1), Y_i(0) \} \mid  (\pi_{11}, \pi_{10}, \pi_{01}, \pi_{00}) &\sim& \mbox{Multinomial}(\pi_{11}, \pi_{10}, \pi_{01}, \pi_{00}), \\
(\pi_{11}, \pi_{10}, \pi_{01}, \pi_{00}) &\sim& \mbox{Dirichlet}(\alpha_{11}, \alpha_{10}, \alpha_{01}, \alpha_{00}).
\end{eqnarray*}
Given the observed data and the parameter, it is straightforward to impute the missing potential outcomes. For example, for units with $W_i = 1$ and $Y_i^\obs = 1$, we draw the missing $Y_i(0)$ from Bernoulli$(\pi_{11}/(\pi_{11} + \pi_{10}))$; for other three types of units, we can similarly draw their missing potential outcomes. After imputing the missing potential outcomes, the counts $(N_{11}, N_{10}, N_{01}, N_{00})$ are known, and we can then draw the parameters from their posterior:
$$(\pi_{11}, \pi_{10}, \pi_{01}, \pi_{00})\mid (N_{11}, N_{10}, N_{01}, N_{00}) \sim \mbox{Dirichlet}(\alpha_{11}+N_{11}, \alpha_{10}+N_{10}, \alpha_{01}+N_{01}, \alpha_{00}+N_{00}).$$
This data augmentation scheme gives us the posterior distributions of $(N_{11}, N_{10}, N_{01}, N_{00})$, which immediately imply the posterior distribution of $\tau^S = (N_{10} - N_{01})/N.$ However, the posterior distribution of $\tau^S$ is sensitive to the choice of the hyperparameters $(\alpha_{11}, \alpha_{10}, \alpha_{01}, \alpha_{00})$, because they determine the \emph{a priori} dependence between the two potential outcomes; see \citet[][Section 3]{ding2014three} for the consequence of using the Jeffreys ``non-informative'' prior: $(\pi_{11}, \pi_{10}, \pi_{01}, \pi_{00})\sim $ Dirichlet$(1/2,1/2,1/2,1/2)$.

Following the second strategy, \citet{ding2016potential} reparametrize the joint distribution of the potential outcomes as $(\pi_{1+}, \pi_{+1}, \gamma)$, where $\pi_{1+} = \pi_{11} + \pi_{10}  = \Pr(  Y_i(1) = 1 ) $ and $\pi_{+1} = \pi_{11} + \pi_{01} = \Pr(   Y_i(0) =1 ) $ govern the marginal distributions, and $\gamma =  \Pr(  Y_i(1)=1\mid Y_i(0)=1 ) /  \Pr(  Y_i(1)=1\mid Y_i(0)=0 )$ governs the association. From $(\pi_{1+}, \pi_{+1}, \gamma)$ we can uniquely determine $(\pi_{11}, \pi_{10}, \pi_{01}, \pi_{00}) $. \citet{ding2016potential} advocated treating $\gamma$ as the sensitivity parameter and obtaining posterior distribution of $\tau^S$ for a fixed $\gamma.$ If we postulate independent priors $\pi_{1+}\sim$ Beta$(\alpha_1,\beta_1)$ and $\pi_{+1}\sim$ Beta$(\alpha_0,\beta_0)$, the posterior distributions are $\pi_{1+}\sim$ Beta$(\alpha_1+n_{11},\beta_1+n_{10})$ and $\pi_{+1}\sim$ Beta$(\alpha_0+n_{01},\beta_0+n_{00})$. After drawing $\pi_{1+}$ and $\pi_{+1}$, we can impute all the missing potential outcomes. For example, for units with $W_i = 1$ and $Y_i^\obs = 1$, we draw $Y_i(0)$ from a Bernoulli$(\gamma\pi_{+1} / (1 - \pi_{+1} + \gamma\pi_{+1}))$; for other three types of units, we can similarly draw their missing potential outcomes. Therefore, we can obtain the posterior distribution of $\tau^S$ for a fixed $\gamma.$ Varying $\gamma$ in a selected range yields a Bayesian sensitivity analysis.$\Box$
\end{example}

\begin{example}\label{example:BPS} \normalfont
[Bayesian post-stratification]
For illustration purpose, we present an example of model-based covariate adjustment in a completely randomized experiment, which is the Bayesian analogue of the Frequentist post-stratification \citep{miratrix2013adjusting}. For each unit $i$, we observed the binary treatment indicator $W_i$, outcome $Y_i$ and a discrete covariate $X_i$ taking values in $\{1,\ldots, K\}.$ Assume the following joint model for the potential outcomes given the covariate:
$$
\begin{pmatrix}
Y_i(1)\\
Y_i(0)
\end{pmatrix} \mid (X_i=k, \theta_{Y|X}  )
\sim  N\left(
\begin{pmatrix}
\mu_{1[k]}\\
\mu_{0[k]}
\end{pmatrix},
\begin{pmatrix}
\sigma_{1[k]}^2 & \rho_{[k]} \sigma_{1[k]} \sigma_{0[k]} \\
\rho_{[k]} \sigma_{1[k]} \sigma_{0[k]} & \sigma_{0[k]}^2
\end{pmatrix}
\right) ,
$$
for $k=1,\ldots, K$ and $i= 1,\ldots, N$, where $\theta_{Y|X}^\text{m} = \{\mu_{1[k]}, \mu_{0[k]},  \sigma_{1[k]}^2,  \sigma_{0[k]}^2 \}_{k=1}^K$ and $\theta_{Y|X}^\text{a} = \{ \rho_{[k]}\}_{k=1}^K$. The observed likelihood factors into two parts: the data in treatment group $\{  (X_i, Y_i^\obs) : W_i = 1 \}$ contribute to the likelihood of $\{\mu_{1[k]},   \sigma_{1[k]}^2 \}_{k=1}^K$, and the data in the control group $\{  (X_i, Y_i^\obs) : W_i = 0 \}$ contribute to the likelihood of $\{ \mu_{0[k]},    \sigma_{0[k]}^2 \}_{k=1}^K$. Importantly, the observed likelihood does not depend on $\theta_{Y|X}^\text{a}.$

Imposing independent priors on all the parameters, we can easily obtain their posterior distributions. For example, the conventional noninformative prior for the Gaussian models yield conjugate posteriors, because within each stratum $X_i=k$ it is a standard Gaussian model. For convenience of description, we define $n_{w[k]}$ as the number of units, and $(\bar{Y}_{w[k]}^\obs,  s_{w[k]}^2)$ as the sample mean and variance of the outcomes under treatment $w$ within stratum $X_i=k$. If $p(\mu_{w[k]}) \propto 1 $ and $p(\sigma^2_{w[k]}) \propto 1/\sigma_{w[k]}^2$ for $w=0,1$ and $k=1,\ldots,K$, then according to \citet[][Section 3.2]{gelman2014bayesian}, the posterior distributions of the parameters are
\begin{eqnarray}
\label{eq::BPS-post}
\sigma_{w[k]}^2 \mid \bO^\obs \sim  (n_{w[k]}  - 1) s_{w[k]}^2 / \chi^2_{n_{w[k]}  - 1},\quad
\mu_{w[k]} \mid  \sigma_{w[k]}^2, \bO^\obs \sim N( \bar{Y}_{w[k]}^\obs,    \sigma_{w[k]}^2/  n_{w[k]}  ).
\end{eqnarray}
For fixed values of $\theta_{Y|X}^\text{a}$ and given each draw of $\theta_{Y|X}^\text{m} $, we can impute the missing potential outcomes as follows: for treated units ($W_i=1$) within stratum $X_i=k$, we draw
$$
Y_i(0)\mid X_i=k, W_i=1,Y_i^\obs, \theta_{Y|X} \sim  N\left(  \mu_{0[k]} + \rho_{[k]} \frac{\sigma_{0[k]} }{\sigma_{1[k]}} (  Y_i^\obs -  \mu_{1[k]} ) ,
\sigma_{0[k]}^2(1-\rho_{[k]}^2)   \right),
$$
and for control units ($W_i=0$) within stratum $X_i=k$, we draw
$$
Y_i(1) \mid X_i=k, W_i=1,Y_i^\obs, \theta_{Y|X}   \sim  N\left(  \mu_{1[k]} + \rho_{[k]} \frac{\sigma_{1[k]} }{\sigma_{0[k]}} (  Y_i^\obs -  \mu_{0[k]} ) ,
\sigma_{1[k]}^2(1-\rho_{[k]}^2)    \right).
$$
Consequently, we obtain the posterior distribution of $\tau^S$. In practice, we suggest varying the $\rho_{[k]}$'s from $0$ to $1$, which correspond to conditionally independent potential outcomes and perfectly correlated potential outcomes. $\Box$
\end{example}

Example \ref{example:BPS} deserves some further discussions. First, for simplicity we can further assume that $\rho_{[k]} = \rho$ and therefore reduce the sensitivity parameters to one. Second, the above strategy works well for small $K$ and large $n_{w[k]}$'s. In particular, the posterior distributions \eqref{eq::BPS-post} are well-defined if $n_{w[k]} \geq 1$, and the marginal posterior mean of $\mu_{w[k]}$ is finite is $n_{w[k]} \geq 3.$ For large $K$ and small $n_{w[k]}$'s, we have a large number of parameters, and therefore need to impose a more sophisticated prior on $\theta_{Y|X}^\text{m}$. Ideally, we want to choose a prior that yields good Frequentist properties under the randomization inference framework, because the original design is a completely randomized experiment. This deserves further research. Third, with a discrete covariate, we can also define the ATE over sub-populations:
$$
\tau^S_k = \frac{1}{N_{[k]}} \sum_{i: X_i=k} \{  Y_i(1) - Y_i(0)   \},
$$
where $N_{[k]} = n_{1[k]} + n_{0[k]}$ is the number of units within covariate stratum $X_i=k$. According to Example \ref{example:BPS}, obtaining the posterior distribution of $\tau^S_k $ is straightforward.

\subsection{Super-population inference}
\label{sec::Bayes_SP}

In general, a super-population causal parameter is a function of the model parameters for the joint distribution of $\{ X_i, Y_i(1), Y_i(0) \}$, written as $\tau = \tau ( \theta_{Y|X} , \theta_X ).$ Therefore, Bayesian inference for super-population causal parameters reduces to obtaining posterior distributions of $\theta_{Y|X}$ and $\theta_X$.
Examples of such causal parameters include the subgroup treatment effect
\begin{equation}\label{est:sub_ace}
\tau^{P}(x) \equiv  \bE\{   Y(1) \mid X=x; \theta_{Y|X}^\text{m} \}  - \bE\{   Y(0) \mid X=x; \theta_{Y|X}^\text{m} \}  ,
\end{equation}
the PATE
\begin{equation}\label{est:ace_sp}
\tau^{P} \equiv  \bE\{ Y_i(1)  - Y_i(0)\}  =  \int  \tau^{P}(x; \theta_{Y|X}^\text{m} )  F_{X}(\d x; \theta_X).
\end{equation}
In most cases, however, we are unwilling to model the possibly multi-dimensional pretreatment covariate $X_i$, and therefore condition on the observed values of the covariates, which is equivalent to replace $\theta_X$ with $ \widehat{\mathbb{F}}_X$, the empirical distribution of $\{X_i \}_{i=1}^N.$ Therefore, most Bayesian causal inferences in fact focus on the conditional ATE
\begin{eqnarray} \label{est:con_ace}
\tau^{\bX} \equiv \int  \tau^{P} (x; \theta_{Y|X}^\text{m} ) \widehat{\mathbb{F}}_X(\d x)
= N^{-1} \sum_{i=1}^N   \tau^{P} (X_i; \theta_{Y|X}^\text{m} ).
\end{eqnarray}
Note that in general $\tau^{\bX} $ is neither the PATE nor the SATE. Because we are often unwilling to model $X$, we omit the discussion of $\tau^{P}$ and focus on $\tau^{\bX} $ in this subsection.

It is worth giving two concrete examples for the causal estimands above. Here we assume that the covariate vector $X_i$ contains an intercept. For a continuous outcome, if $Y_i(w)\mid X_i\sim N(\beta_w' X_i, \sigma_w^2)$ for $w=1,0$, then
$
\tau^P(x) = (\beta_1 - \beta_0)' x
$
and
$
\tau^\bX = N^{-1}\sum_{i=1}^N (\beta_1 - \beta_0)' X_i =(\beta_1 - \beta_0)' \bar{X} .
$
For a binary outcome, if $\Pr(Y_i(w)=1\mid X_i) = F(\beta_w' X_i)$ for $w=1,0$ with $F(\cdot)$ being the link function (for example, the standard normal distribution $F(\cdot) = \Phi(\cdot)$ gives the probit model), then
$
\tau^P(x) = F(\beta_1'x) - F(\beta_0'x)
$
and
$
\tau^\bX = N^{-1}\sum_{i=1}^N \{ F(\beta_1'X_i) - F(\beta_0'X_i) \}.
$

A salient feature of the estimands in Equation \eqref{est:sub_ace}--\eqref{est:con_ace} is that they depend only on $\theta_{Y|X}^\text{m}$ but not on $\theta_{Y|X}^\text{a}$. For this type of estimands, we do not need to impose a joint model of $\{ Y_i(1), Y_i(0) \}$ given $X_i$, because the likelihood and thus the posterior distribution of $\theta_{Y|X}^\text{m}$ requires only the specification of the marginal distributions of $Y_i(w)$ given $X_i$ for  $w=0,1$.  Bayesian inference for estimands \eqref{est:sub_ace}--\eqref{est:con_ace} is straightforward once we obtain the posterior distributions of the parameter $\theta_{Y|X}^\text{m}$ following the second strategy in Section \ref{sec:Bayes_FP}.

Moreover, we can infer parameters that depend on the joint distribution of the potential outcomes. For example, we can consider $\delta_1 = \Pr(Y_i(1)\geq Y_i(0))$ and its conditional version
$
\delta_1^\bX =  N^{-1} \sum_{i=1}^N \delta_1(X_i),
$
where
$$
\delta_1(x) =  \Pr(Y_i(1)>Y_i(0)\mid X_i=x, \theta_{Y|X}^\text{m},  \theta_{Y|X}^\text{a}).
$$
We give an example to illustrate inferring $\delta_1^\bX$ under a Gaussian linear model as in \citet{Heckman14}.

\begin{example}\label{eg:delta2-gaussian} \normalfont [Gaussian linear model]
Assume
\begin{eqnarray}
\begin{pmatrix}
Y_i(1)\\
Y_i(0)
\end{pmatrix} \mid (X_i, \theta_{Y|X}  )
\sim  N\left(
\begin{pmatrix}
\beta_1' X_i\\
\beta_0' X_i
\end{pmatrix},
\begin{pmatrix}
\sigma_1^2 & \rho \sigma_1 \sigma_0 \\
\rho \sigma_1 \sigma_0 & \sigma_0^2
\end{pmatrix}
\right) ,\quad (i= 1,\ldots, N),
\label{eq::jointnormal}
\end{eqnarray}
where $\theta_{Y|X}^\text{m} = (\beta_1, \beta_0, \sigma_1^2, \sigma_0^2)$ and $\theta_{Y|X}^\text{a} = \rho$. The model \eqref{eq::jointnormal} implies
\begin{eqnarray}
\label{eq::marginal-gaussian}
Y_i(w)  \mid X_i, \theta_{Y|X}^\text{m} \sim N(\beta_w' X_i, \sigma_w^2  ),\quad (w=0,1)
\end{eqnarray}
and
\begin{eqnarray}
\label{eq::difference-gaussian}
Y_i(1) - Y_i(0)\mid X_i, \theta_{Y|X} \sim N(  (\beta_1 - \beta_0)'X_i,  \sigma_1^2 + \sigma_0^2 - 2\rho \sigma_1 \sigma_0 ).
\end{eqnarray}
Therefore, we can derive from \eqref{eq::difference-gaussian} that
$$
\delta_1^\bX = \frac{1}{N} \sum_{i=1}^N  \Phi\left\{    \frac{  (\beta_1 - \beta_0)' X_i  }{  ( \sigma_1^2 + \sigma_0^2 - 2\rho \sigma_1 \sigma_0 )^{1/2} }  \right\}.
$$
Based on the marginal model \eqref{eq::marginal-gaussian} we can obtain the posterior distribution of $\theta_{Y|X}^\text{m}$, for example, via standard Gaussian posterior calculations. Since the observed data do not contain any information about $\rho$, we can vary it from $0$ to $1$, and obtain the posterior distribution of $\delta_1^\bX $ for each fixed $\rho.$
\end{example}

The Bayesian framework offers a unified and flexible approach to inferring causal parameters in complex settings. Indeed, there is a rapidly growing literature in applying advanced Bayesian models and methods---such as Bayesian nonparametric methods, Bayesian model selection and model averaging---to causal inference. A comprehensive review of the Bayesian approach to causal inference is beyond the scope of this paper, we leave that to another review article \citep{li2018bayesian}.

\section{The Fisherian Randomization Perspective}
\label{sec:Fisherian}

The Fisherian randomization perspective to causal inference focuses on p-values under null hypotheses obtained by comparing the observed values of test statistics with their randomization distribution \citep{fisher1935design}. Broadly speaking, the Fisherian perspective belongs to the Frequentist perspective because the p-value is commonly viewed as a Frequentist notion. For its special emphases and historical reasons, we classify the Fisherian perspective as a separate inferential framework. This perspective is rather unique to causal inference because standard missing data problems do not involve comparisons of different treatment groups. The Fisherian framework is also closely related to the Bayesian perspective in that a key step is to impute all the missing potential outcomes.

Because the Fisherian framework has been largely developed in randomized studies, we will focus on the finite-sample inference.

\subsection{Fisher randomization test of the sharp null hypothesis}
\label{subsec::sharpnull-ppc}

The Fisherian randomization test considers a finite population of $N$ units with $N_1$ receiving treatment and $N_0$ receiving control. For the ease of exposition, we discuss the case with a known propensity score $\Pr(\bW \mid  \bX)$ such as in (stratified) completely randomized experiments. We first present a general form of the Fisher randomization test for a sharp null hypothesis that $Y_i(1) - Y_i(0) = \tau_i$ with a known constant vector $\bm{\tau} = (\tau_1, \ldots, \tau_N)$. The null hypothesis is sharp because it allows for imputing all the missing potential outcomes: for a unit with $W_i = 1$, we have $Y_i(1) = Y_i^{\text{obs}} $ and $Y_i(0)  = Y_i^{\text{mis}} =  Y_i^{\text{obs}} - \tau_i$; for a unit with $W_i = 0$, we have $  Y_i(0) =  Y_i^{\text{obs}}$ and $ Y_i(1)  = Y_i^{\text{mis}} =  Y_i^{\text{obs}} +\tau_i$. Originally, \citet{fisher1935design} considered the sharp null $H_0^\#$ with $\tau_i = 0$ for all $i=1,\ldots, N$, or equivalently $\bY(1) = \bY(0) = \bY^\obs .$

In general, a test statistic $t(\bW, \bY(1), \bY(0), \bX)$ is a function of the treatment assignment vector, potential outcomes and pretreatment covariates. For example, it can be the difference-in-means of the outcomes, $N_1^{-1} \sum_{i=1}^N W_i Y_i(1) - N_0^{-1} \sum_{i=1}^N (1-W_i ) Y_i(0)$, or the difference-in-means of the residuals from linear regressions of $\bY(1)$ and $\bY(0)$ on $\bX$, $N_1^{-1} \sum_{i=1}^N W_i \epsilon_{i1} - N_0^{-1} \sum_{i=1}^N (1-W_i ) \epsilon_{i0}$, where $\bm{\epsilon}_1$ and $\bm{\epsilon}_0$ consist of the residuals from the linear regressions of $\bY(1)$ and $\bY(0)$ on $\bX$ without an intercept, respectively \citep{tukey1993tightening, rosenbaum2002covariance}. Because $\bX$ is fixed and $\{  \bY(1), \bY(0) \} $ are known under the null, the distribution of $t(\bW, \bY(1), \bY(0), \bX)$ is determined by the only random component $\bW$. Therefore, we can calculate or simulate the randomization distribution of the test statistic according to $\Pr(\bW \mid \bX)$, and then obtain the p-value $\Pr(\widetilde{t} \geq t)$, defined as the tail probability of the test statistic $t$ with respect to its randomization distribution $\widetilde{t}$.

\subsection{Fisherian p-value as a posterior predictive p-value}\label{sec:ppp}

We first review the posterior predictive p-values. Although general forms exist \citep{Gelman96}, we use \citet{meng1994posterior}'s formulation of the posterior predictive p-value (ppp) and tailor it to the finite-sample causal inference. Recall that $\bO^\obs$ and $\bO^\mis$ denote the observed and missing data, respectively. Consider testing a null hypothesis $H_0.$

If $\bO^\mis$ were known, then we can choose a test statistic $t(\bO^\obs)$ to measure a deviation from $H_0$, and obtain the p-value $\text{p}(\bO^\mis)$. We highlight the dependence of the p-value on $\bO^\mis$, because the distribution of $t(\bO^\obs)$ depends on $\bO^\mis$ in general. The ppp with missing potential outcomes is defined as the mean of $\text{p}(\bO^\mis)$ over the posterior distribution of $\bO^\mis$ given the observed data $\bO^\obs$, $\Pi(  \bO^\obs  \mid \bO^\obs  )$,
\begin{equation}\label{eq:ppp}
\text{ppp} = \int \text{p}(\bO^\mis )   \Pi(  \bO^\mis   \mid \bO^\obs  ).
\end{equation}
In order to obtain the posterior distribution of $\bO^\mis$, we often need to invoke a model for the complete data $\Pr(\bO^\obs, \bO^\mis \mid \theta)$ and a prior $p(\theta)$. The joint posterior of $(\bO^\mis, \theta)$ can be obtained from
$$
\Pr( \bO^\mis, \theta \mid \bO^\obs  ) \propto \Pr(\bO^\obs, \bO^\mis \mid \theta) p(\theta),
$$
and the posterior $\Pi(  \bO^\mis  \mid \bO^\obs  )$ can be obtained by marginalizing over $\theta.$

The above formulation is generic and applicable to general finite-sample causal inference. Below we discuss an application, using the Fisher randomization test to obtain $\text{p}(\bO^\mis)$ and using a Bayesian model to obtain the posterior distribution of the missing potential outcomes. Therefore, the ppp discussed in this subsection has both Bayesian and Fisherian favors.

The pure randomization-based test was often viewed as a model-free robust procedure that had no connection to the Bayesian inference. However, \citet{Rubin84} first interpreted the Fisher randomization test as a Bayesian posterior predictive check. We assume that $\Pr(\bW\mid  \bX)$ has a known probabilistic law as in randomized experiments. With a known unit-level treatment effect vector $\bm{\tau} $,  the posterior distribution $\Pi(  \bO^\mis   \mid \bO^\obs  )$ contains point masses at the imputed values of the missing potential outcomes. Therefore, the posterior predictive replicates can be obtained by drawing $\bW$ according to its probabilistic law without changing the values of $\{   \bX, \bY(1), \bY(0)  \} $, and the ppp using statistic $t(\bW, \bY(1), \bY(0), \bX)$ is identical to that obtained from the Fisher randomization test in Section \ref{subsec::sharpnull-ppc}.

\subsection{Non-sharp null hypotheses} \label{sec::fisher-nonsharp}

Although the Bayesian interpretation of the Fisher randomization test in Section \ref{sec:ppp} is ``intellectually pleasing'' \citep{Rubin05}, its real advantages are not obvious in the case of a sharp null hypothesis. When the null hypothesis is not sharp, it is often challenging to conduct exact randomization tests without sacrificing statistical power \citep{nolen2011randomization, ding2015randomization}. Fortunately, we can still obtain ppp's in these cases, because we can first obtain p-values for fixed values of the unknown missing potential outcomes, and then average the p-values over the posterior distributions of these unknown potential outcomes. We give two examples to illustrate this general approach.

\begin{example}\normalfont
[Testing treatment effect variation] We are interested in testing whether the unit-level treatment effects are constant \citep{ding2015randomization}, a model that is often assumed in randomization-based causal inference \citep{RosenbaumBook02}:
$$
H_0^C:  Y_i(1) - Y_i(0 ) = \tau  ,\quad (\text{for some } \tau \text{ and for all } i=1, \ldots, N).
$$
Because the null hypothesis depends on an unknown parameter $\tau$, it is not sharp and thus the missing potential outcomes cannot be imputed as in Section \ref{subsec::sharpnull-ppc}. However, for a fixed value of $\tau$, the null hypothesis that  $Y_i(1) - Y_i(0 ) = \tau$ for all unit $i$ is sharp, and all the missing potential outcomes can be imputed: for a unit with $W_i = 1$, we have $ Y_i(1) = Y_i^{\text{obs}} $ and $Y_i(0) = Y_i^{\text{mis}} = Y_i^{\text{obs}} - \tau$; for a unit with $W_i = 0$, we have $ Y_i(0) = Y_i^{\text{obs}} $ and $Y_i(1) = Y_i^{\text{mis}} = Y_i^{\text{obs}} + \tau$.
\citet{ding2015randomization} suggested using the following shifted Kolmogorov--Smirnov statistic for $H_0^C$:
$$
t_{\text{SKS}} (\bW, \bY(1), \bY(0), \bX) =  \max_y \Big |
N_1^{-1} \sum_{W_i=1}   I(Y_i^{\text{obs}}  - \widehat{\tau} \leq y ) - N_0^{-1}  \sum_{W_i=0}  I(Y_i^{\text{obs}}  \leq y )
\Big | ,
$$
where $\widehat{\tau}$ is the difference in means of the outcomes in treatment and control groups. For a fixed value of $\tau$, we can first impute all the missing potential outcomes, and then obtain the p-value
$
\text{p}(\tau) = \Pr( \widetilde{t}_{\text{SKS}} \geq t_{\text{SKS}}  \mid \tau  )
$
using the Fisher randomization test, where $\widetilde{t}_{\text{SKS}}$ represents the randomization distribution of the test statistic $t_{\text{SKS}}$ that can be simulated by Monte Carlo. Note that in this example, because under $H_0^C$ there is a one-to-one map between $\tau$ and $\bO^\mis = \bY^\mis$, we use the simple notation $\text{p}(\tau)$ for $\text{p}(\bO^\mis ).$

However, we do not know $\tau$ and need to obtain its posterior distribution. Assume that the first part of the joint model $\Pr(\bW\mid \bX)$ is known as in (stratified) completely randomized experiments. We need only to model the second part $ \Pr(  \bY(1), \bY(0)\mid  \bX, \tau, \theta )  = \Pr(  \bY(0)\mid  \theta)  \times I\{  \bY(1) = \bY(0)  + \tau \bm{1}_N \}$ and impose priors on $(\tau, \theta)$. For notational simplicity, we illustrate the idea with a Normal linear model $\bY(0) = \bX \beta + \bm{\varepsilon}$ with an intercept, where $\bm{\varepsilon} \sim N(\bm{0}, \sigma^2 \bm{I}_N)$ and $\theta = (\beta, \sigma^2)$. As a result, the observed outcomes follow $\bY^{\text{obs}}  =  \tau \bW +   \bX \beta + \bm{\varepsilon}$, and under the usual normal and inverse-$\chi^2$ priors for $(\tau, \beta, \sigma^2)$, the posterior of $\tau$ can be easily obtained in close form \citep[][Chapter 14]{gelman2014bayesian}. The final ppp-value is obtained by averaging p$(\tau)$ over the posterior distribution $\Pi(\tau \mid  \bO^\obs)$, which is
$
\text{ppp}(H_0^C) =  \int p(\tau)  \Pi( \d \tau \mid  \bO^\obs  ) .
$

\citet{ding2015randomization} also discussed several alternative test statistics. In practice, we may also want to impose more flexible outcome models beyond the Normal linear model, but the essence of the ppp remains the same.
$\Box$
\end{example}

\begin{example}\normalfont \label{ex::noncompliance-fisher}
[Testing treatment effects among compliers]
We revisit the principal stratification approach to noncompliance (Section \ref{sec::freq-PS}) under the Fisherian perspective. Recall that for unit $i$, let $X_i$, $Z_i$, $W_i(1)$, $W_i(0)$, $Y_i(1)$ and $Y_i(0)$ be covariate, treatment assignment, potential values of the actual treatment and the outcome, respectively. Recall that $U_i = \{ W_i(1), W_i(0) \}$ is the latent principal stratum. \citet{rubin1998more} and \citet{RosenbaumBook02} considered testing the following null hypothesis for the compliers:
$$
H_0^\co : Y_i(1) = Y_i(0), \quad (\text{for } U_i = \co, ~ i=1,\ldots, N).
$$
Assuming monotonicity and exclusion restriction, this null hypothesis is equivalent to Fisher's sharp null hypothesis $H_0^\#$ in Section \ref{subsec::sharpnull-ppc} that the treatment does not affect the outcomes of any units. This equivalence implies that we can simply conduct the usual Fisher randomization test discussed in Section \ref{subsec::sharpnull-ppc} for $H_0^\co$ based on the data $(\bX, \bZ, \bY^\obs)$, ignoring the information of noncompliance $\bW^\obs$. However, this exact randomization test does not make full use of the observed data and consequently can lose power in some cases.

Define $( \bar{Y}_1^\obs ,  \bar{Y}_0^\obs )  $ and $  (\bar{W}_1^\obs , \bar{W}_0^\obs  )$ as the observed means of the treatments received and outcomes under treatment and control assignments. Two commonly-used estimators for the CACE are \emph{(i)} the Wald estimator $ t_{\text{Wald}}(  \bZ,  \bW(1), \bW(0), \bY(1), \bY(0)  , \bX)  =( \bar{Y}_1 - \bar{Y}_0 )  / (\bar{W}_1 - \bar{W}_0)$ \citep{Angrist96}, and \emph{(ii)} the 2SLS estimator $ t_{\text{TSLS}}(  \bZ,  \bW(1), \bW(0), \bY(1), \bY(0)  , \bX)$ that is the coefficient of $\bW^\obs $ in the regression $\bY^\obs \sim    \bW^\obs + \bX $ using $\bZ$ as an instrument for $\bW^\obs$ \citep{angrist2008mostly}. It is intuitive to use one of them as a test statistic in the randomization test. Unfortunately, these test statistics and their null distributions depend on the unknown potential values of the treatment received $\bW^\mis$: for unit with $Z_i=1$, we have $W_i^\mis = W_i(0)$, and for units with $Z_i=0$, we have $W_i^\mis = W_i(1)$. If $\bW^\mis$ were known, then we can simulate the distribution of the test statistic (e.g., $t_{\text{Wald}}$ or $ t_{\text{TSLS}}$) to obtain the p-value p$(\bW^\mis) = \Pr( \widetilde{t} \geq t \mid  \bW^\mis )$, where $\widetilde{t}$ is the posterior replicates of the test statistic $t$. The final ppp-value is the posterior mean of $\text{p}(\bW^\mis)$ given the data, which is
$
\text{ppp}(H_0^\co   ) =  \int   \text{p}(\bW^\mis)   \Pi(  \d \bW^\mis \mid  \bO^\obs  ).
$

Operationally, the key is to obtain the posterior draws of $ \bW^\mis$. Under $\bY(1) = \bY(0) = \bY^\obs$, the observed outcomes $\bY^\obs$ can be viewed as {\it pretreatment covariates} under monotonicity, exclusion restriction and $H_0^\co$, and therefore we do not need to model the outcome. Assume that first part of the joint model $\Pr(\bZ \mid  \bX, \bU,  \bY^\obs) = \Pr(\bZ \mid  \bX ) $ is known, as often guaranteed by the design of experiments. Therefore, we need only to model the second part $\Pr(\bU \mid  \bX, \bY^\obs, \theta)$ and impose a prior $p(\theta)$. Assuming exchangeability of the units, we define $\pi_{u}(X_i, Y_i^\obs, \theta) = \Pr( U_i = u \mid    X_i, Y_i^\obs, \theta)$ for $u = \co, \nt$ and $\at$, which, for example, can be modeled as Multinomial Logistic. We can obtain the posterior distribution $\Pi(  \bW^\mis \mid  \bO^\obs  )$ as a byproduct of the iteration of the data augmentation algorithm for sampling $(\bW^\mis, \theta)$, by discarding the posterior draws of $\theta$. In particular, given $\bW^\mis $, we know $\bU$ and therefore we can use $(\bU, \bX, \bY^\obs)$ to obtain a posterior draw of $\theta$, for example, according to the posterior distribution of a Multinomial Logistic model. Given $\theta$, we can impute $\bW^\mis $ as follows: for a unit with $(Z_i=1,W_i^\obs=1)$, draw $ W_i(0)$ from Bernoulli with probability $ \pi_{\at}(X_i, Y_i^\obs, \theta) /  \{ \pi_{\at}(X_i, Y_i^\obs, \theta)  + \pi_{\co}(X_i, Y_i^\obs, \theta) \} $; for a unit with $(Z_i=1,W_i^\obs=0)$, set $W_i(0) = 0$; for a unit with $(Z_i=0,W_i^\obs=1)$, set $W_i(1) = 1$; for a unit with $(Z_i=0,W_i^\obs=0)$, draw $ W_i(1)$ from Bernoulli with probability $ \pi_{\co}(X_i, Y_i^\obs, \theta) /  \{ \pi_{\co}(X_i, Y_i^\obs, \theta)  + \pi_{\nt}(X_i, Y_i^\obs, \theta) \} $. \citet{rubin1998more} described this strategy for the one-sided noncompliance with $W_i(0)=0$ for all units under a completely randomized experiment; simulation studies for the case with a binary outcome without $\bX$, showed meaningful power gains in some scenarios.
$\Box$
\end{example}

\subsection{Extension}\label{sec::remark-fisher}

In Section \ref{sec::fisher-nonsharp}, we focused on obtaining the p-value based on a test statistic $t(\bO^\obs)$, a function of the observed data only. In general, we can use a {\it discrepancy variable} $t(\bO^\obs, \bO^\mis)$, which can be a function of both the observed and missing data, to obtain the p-value \citep{meng1994posterior, Gelman96}. In both cases, the p-value is a function of $\bO^\mis.$

The above discussion of the Fisher randomization inference applies naturally to randomized studies in which $\Pr(\bW\mid \bX)$ is known and determined by the designers of the experiments. Complications arise in observational studies. If Assumption \ref{ass:unconfounded} holds and $X$ is discrete, then we can use the estimated propensity score $\widehat{\Pr}(\bW\mid \bX)$ to simulate the treatment $\bW$, or equivalently conduct conditional randomization test as if the data come from a stratified completely randomized experiment. With continuous or multi-dimensional $X$, we need to model $\Pr(\bW\mid \bX)$. \citet{rosenbaum1984conditional} proposed a conditional randomization test given the sufficient statistics of a logistic model for $\Pr(\bW\mid \bX)$. \citet{Rubin07, Rubin08} and \citet{ImbensRubin15} suggested stratifying on the estimated propensity scores to approximate completely randomized experiments within strata, possibly followed by a Fisherian analysis.

Although we focused on obtaining p-values from Fisher randomization tests, we can invert a sequence of tests to obtain confidence sets of parameters of interest. With a few exceptions, this is often conducted under an additional assumption of constant treatment effect \citep{RosenbaumBook02}. Moreover, with non-sharp null hypotheses, simultaneously imputing missing potential outcomes and inverting a sequence of tests seems inferior to both the Frequentist and Bayesian perspectives discussed before, which focus on point and interval estimation directly.

Moreover, it is important to evaluate the frequency properties of the ppp's. Under the classical Frequentist evaluation in which the parameters are fixed constants, the Fisherian p-values are {\it exact} when the null hypotheses are sharp, but they are often conservative when the null hypotheses are not sharp \citep{robins2000asymptotic}. However, alternative frequency evaluations often give different conclusions.
\citet{meng1994posterior} considered the {\it prior predictive evaluation}, in which the model parameters are generated from proper prior distributions and the data are then generated conditional on the model parameters. Under \citet{meng1994posterior}'s scheme, some ppp's may be anti-conservative. \citet{rubin1998more} considered the {\it posterior predictive evaluation}, in which the model parameters and missing data are generated conditional on the observed data and the null hypotheses. Under \citet{rubin1998more}'s scheme, the ppp's have exact frequency properties.

As a final remark on the Fisherian perspective, neither did Fisher formally use potential outcomes nor did he agree with using Bayes' theorem for statistical inference \citep{fisher1935design}.  \citet{Rubin80} reformulated the randomization test using \citet{Neyman23}'s potential outcomes notation, and extended it by combining p-values with Bayesian techniques. Alternatively, assuming that the units are independent and identically draws from a super population, \citet{hoeffding1952large} and \citet{chung2013exact} stated the ``sharp'' null hypothesis as $Y_i(1) \sim Y_i(0)$, i,e., the treated and control potential outcomes have the same distribution, and interpreted randomization tests as {\it permutation tests}. This interpretation works well under the sharp null hypotheses for completely randomized experiments, but is less straightforward for non-sharp null hypotheses with nuisance parameters. Super-population version of the analysis in Section \ref{sec::fisher-nonsharp} remains an open question.

\section{Discussion} \label{sec:discussion}

We have reviewed a wide range of causal inference methods that have analogues in missing data analysis under three modes of inference. Although a comprehensive review of all relevant topics is beyond the scope of this paper given the vast literature in both causal inference and missing data, we regard the following important areas merit further attention.

\subsection{Partially identified parameters and bounds} \label{sec:bounds}

In the Bayesian perspective, we commented on the identifiability issue due to the fundamental problem of causal inference and recommended to use a transparent parametrization strategy (see Section \ref{sec:Bayes_FP}), which also applies to the Fisherian perspective. We now comment more on the identifiability issue in the Frequentist perspective, which often involves bounding the parameter of interest by the distributions of the observables. \citet{cochran1953sampling} derived bounds for non-ignorable missing data in surveys, and \citet{manski1990nonparametric} obtained more fruitful results for both missing data and causal inference problems.

The first important class of partially identified parameters depend on the association between the potential outcomes. For example, the parameters
\begin{eqnarray*}
\delta_1 =  \Pr\{ Y(1) \geq Y(0) \},\quad \delta_2 = \Pr\{  Y(1) > Y(0)   \}
\end{eqnarray*}
measure the probability that the treatment is not worse than the control and the probability that the treatment is better than the control, respectively. \citet{lu2015treatment} emphasize that for ordinal outcomes, $\delta_1$ and $\delta_2$ are well defined even though $\tau_i = Y_i(1) - Y_i(0)$ is not. In general, we are interested in the distribution of the treatment effect $\Delta(c) = \Pr \{ Y_i(1) - Y_i(0) \leq  c \}$. Without imposing further assumptions, we can only bound parameters such as $\delta_1$, $\delta_2$ and $ \Delta(c) $ by the marginal distributions $    \Pr\{ Y_i(1) \leq y_1  \}$ and $   \Pr\{ Y_i(0) \leq y_0  \}$.

Inferring such parameters is arguably more challenging than most standard missing data problems. It is also an example where some theoretical development in causal inference lends to research on missing data. Specifically, a small but growing literature is on the so-called ``misaligned missing data'' problem \citep[e.g.,][]{ding2016algorithm}, where some variables cannot be observed simultaneously, similar to the joint potential outcomes $\bY(1)$ and $\bY(0)$. \citet{ding2016algorithm} made the connection and borrowed some results from causal inference to address the problem \citep{fanpartial}.

Inference involving simultaneous counterfactual potential outcomes is controversial for some researchers who were only willing to model observables. For example, \citet{dawid2000causal} advocated ``causal inference without counterfactuals" through a decision theoretic perspective. Interestingly, despite his original critique, \citet{dawid2017probability} invoked counterfactuals to define the ``probability of causation (PC).'' For example, if a unit responds to the treatment, then
$$
\text{PC} = \Pr(Y_i(0)=0\mid W_i=1, Y_i(1)=1)
$$
is the probability that the treatment is effective. The parameter PC obviously involves the joint values of the counterfactuals. Indeed, such joint counterfactuals are often unavoidable in defining certain causal estimands, such as $\delta_1$, $\delta_2$ and $ \Delta(c) $. \citet{Heckman14} provided more discussions on this issue.

Another important class of partially identified parameters arises from principal stratification. If we do not invoke monotonicity, exclusion restriction or parametric models, in general we cannot identify the means of the potential outcomes within the latent principal strata. \citet{Zhang03} derived large-sample bounds of the causal parameters based on the observed data, followed by \citet{cheng2006bounds}, \citet{grilli2008nonparametric}, \citet{imai2008sharp}, among others.

Unfortunately, the bounds are often too wide to be useful in practice, and additional information is often required to sharpen them \citep{mattei2011augmented, yang2016using}. It is also nontrivial to construct confidence intervals for the bounds or for the parameters themselves, because the bounds often correspond to non-smooth operations of the observed data distribution and therefore the standard delta-method or the bootstrap may not apply \citep{andrews2000inconsistency, fan2010sharp}.

\subsection{Nonignorable assignment mechanisms and sensitivity analysis}

Our discussion is limited to ignorable (or unconfounded) assignment mechanisms. However, in observational studies the assignment mechanism is generally unknown, and it is the norm rather exception that the untestable unconfoundedness assumption \eqref{am:unconfounded} is violated to a certain degree. The critical reliance on unconfoundedness in causal inference is similar to the dependence on the vulnerable MAR assumption in missing data analysis. In causal inference, the standard approach is to conduct sensitivity analysis around unconfoundedness. The main idea is to examine the causal estimates from the same inferential procedure given an unmeasured covariate whose confounding effects (i.e., degree of violation to unconfoundedness) are encoded as sensitivity parameters and are varied within a realistic range by the analyst. \citet{Cornfield::1959} first started this school of thoughts, whose method was directly extended in \citet{ding2016sensitivity}. \citet{Rosenbaum83sensitivity} broadened this approach by parametrically modeling the distributions of the outcome and the treatment conditional on the unmeasured confounder; semi- and nonparametric versions of this method have been developed \citep[e.g.][]{rosenbaum1987sensitivity, imbens2003sensitivity, Ichino::2008, robins2000sensitivity}, and extension to principal stratification is also available \citep[e.g.][]{schwartz2012sensitivity, mercatanti2017debit, ding2016principal}.

Sensitivity analysis is also extensively conducted in the context of nonignorable missing data, but with different emphasis and implementation than that in causal inference. Specifically, the two standard models for nonignorable missing data are the \emph{selection} models and the \emph{pattern-mixture} models \citep[][Chapter 15]{little2002missing}; in both models, the missing data mechanism is directly modeled and estimated from the data, together with the outcome model. Because sensitivity to model specification is a serious scientific problem for both selection and pattern-mixture models, in real applications it is prudent to consider a variety of missing data models, rather than to rely exclusively on one model. One attractive direction for both causal inference and missing data with nonignorable assignment/missing data mechanisms is to utilize external data sources to augment the analysis, which requires considerable efforts in the study design.

\subsection{Unintentional missing data}

We have mainly focused on the intentional missing data (i.e. missing potential outcomes) in causal inference. Unintentional missing data are prevalent in observational studies and there is a growing literature on drawing causal inference from incomplete data. A straightforward approach consists of two independent steps: first impute the missing data and then draw causal inferences from the imputed complete data \citep[e.g.][]{mitra2011estimating}. However, how the missing values are imputed may have a nontrivial impact on the subsequent causal analysis \citep[e.g.][]{mitra2016comparison}. In particular, missing data in covariates \citep{rosenbaum1984reducing, ding2014identifiability, yang2017nonparametric}, treatment \citep{molinari2010missing, mebane2013causal, zhang2016causal}, and outcomes \citep{Frangakis99, chen2009identifiability, MatteiMealliPacini2014} are often of different nature and may require different handling. The problem is even more challenging in complex settings such as post-treatment variables \citep[e.g.][]{mercatanti2004analyzing, mealli2004analyzing}. More research in understanding such an impact would be valuable to practice.

\subsection{Software}
Open-source and user-friendly software packages are crucial for bridging theory and practice, and deserve much effort and investment from methodologists. In statistics, authors are increasingly releasing R packages implementing their methods. This is particularly important for causal inference methods, which are widely used in many substantive disciplines including medicine, policy, psychology, social sciences and others. We list a few most popular packages here: \texttt{twang} \citep{twang} provides functions for propensity score estimating (via generalized boosted models) and weighting methods; packages \texttt{Matching} \citep{Matching} and \texttt{MatchIt} \citep{MatchIt} provide functions for matching methods. These packages can be used in combination with other R packages, e.g., for hierarchial models (\texttt{lme4}), Bayesian modeling (\texttt{rjags}), to apply to more complex problems. Besides R, \texttt{Stan} \citep{Stan} is new open-source probabilistic programming language particulary suitable for advanced statistical modeling and computation; \texttt{Stan} has attracted much interest from both academia and industry recently and can be used for implementing many causal inference methods discussed in this review.

%

%
%

\bibliographystyle{natbib}
\bibliography{missingCI_review}

\end{document}